\documentclass[prd,eqsecnum,nofootinbib,twocolumn,letterpaper]{revtex4}

\usepackage{bm,graphicx,mathrsfs}
\usepackage{amssymb,amsmath}

\usepackage[colorlinks=true,citecolor=blue,bookmarks=true,hyperfootnotes=false]{hyperref}

\newcommand{\CKY}{\mathcal{K}}

\newcommand{\calZ}{\mathcal{Z}}
\newcommand{\calP}{\mathscr{P}}
\newcommand{\calM}{m}

\newcommand{\PSi}{\Psi}
\newcommand{\DeltaK}{\Delta_K}
\newcommand{\SigmaK}{\Sigma_K}
\newcommand{\rt}{r_t}
\newcommand{\Jc}{\tilde{J} }

\newcommand{\jS}{ \mathscr{J} }

\newcommand{\Lie}{ \mathscr{L} }

\newcommand{\rc}{ r_c }

\renewcommand{\Re}{\operatorname{Re}}
\renewcommand{\Im}{\operatorname{Im}}

\begin{document}
\title{Extended-body motion in black hole spacetimes: What is possible?}

\author{Abraham I. Harte}

\address{Centre for Astrophysics and Relativity, School of Mathematical Sciences
	\\
	Dublin City University, Glasnevin, Dublin 9, Ireland}

\begin{abstract}
    Free-fall is only approximately universal in general relativity: Different extended bodies can fall in different ways, depending on their internal dynamics. Nevertheless, certain aspects of free-fall are independent of those dynamics. This paper derives universal constraints on extended-body motion which hold in all vacuum type D spacetimes. Working in the quadrupole approximation, we show that in addition to the (previously-known) constraints imposed by Killing vectors, two components of the gravitational torque must vanish. Furthermore, of the ten components of a body's quadrupole moment, four are found to be irrelevant, two can affect only the force, and the remaining four can affect both forces and torques. As an application, we consider the capabilities of a hypothetical spacecraft which controls its motion by controlling its internal structure. In the Schwarzschild spacetime, such a spacecraft can control its mass, and by doing so, it can stabilize unstable orbits, escape from bound orbits, and more---all without a rocket. 
\end{abstract}

\maketitle

\vskip 2pc

\section{Introduction}

A geodesic can be uniquely fixed by specifying an initial position and an initial velocity. To the extent that bodies in general relativity move on geodesics, free-fall is therefore universal: All objects with the same initial position and the same initial velocity have the same past and future. However, it is only approximately true that freely-falling objects move on geodesics. More completely, the laws of motion depend on a body's spin as well as its quadrupole and higher moments. And because the time dependence of the quadrupole and higher moments depends on a body's small-scale features, different bodies can fall in different ways. This paper explores which aspects of extended-body motion are nevertheless universal, obtaining constraints which hold for all possible bodies in all vacuum type D spacetimes (including Schwarzschild and Kerr). While our focus is on extended test bodies in the quadrupole approximation, some results also hold at higher multipole orders. 

Much of the prior literature on quadrupole effects in general relativity has been motivated by astrophysics. It has therefore focused on  bodies which have special types of moments and which move in Schwarzschild or Kerr backgrounds \cite{Bini2008, Steinhoff2012, Bini2013, Bini2014, Bini2014(2), Bini2015, Bini2015(2), Han, SteinhoffReview}. In some cases, quadrupole moments were assumed to be permanent and to evolve quasi-rigidly \cite{Bini2008, Bini2013, Bini2014, Bini2014(2)}. In others, moments were assumed to be induced---determined either by a body's spin \cite{Steinhoff2012, Bini2015(2), Han, SteinhoffReview} or by an external tidal field \cite{Steinhoff2012, SteinhoffReview}. While these are reasonable models for self-gravitating astrophysical objects, they are not the only possibilities allowed by the laws of physics.

Here, we allow for more general moments and more general metrics. There are two reasons for this. First, weakening assumptions clarifies the underlying theory. It allows us to see that the Petrov type of a spacetime can be used to simplify the laws of motion, and that in type D spacetimes, conformal Killing-Yano tensors play an important role. Our second reason for generalizing previous work is that there may be interesting systems whose quadrupole moments differ from those which have  already been considered. It is possible that some astrophysical systems behave unexpectedly, and it is important to understand the space of possibilities.

Considering more general moments also allows us to address questions which are not motivated by astrophysics. For example, can a spacecraft control its motion simply by controlling its internal structure? Indeed it can. Even in Newtonian gravity, a spacecraft which manipulates its shape can modulate the forces and torques which act upon it. While these modulations may be small, their effects can grow arbitrarily large  \cite{Beletsky, GratusTucker, MartinezSanchez, HarteNewtonian}. This phenomenon requires that a body's moments be neither quasi-rigid nor induced. By investigating motion in a more general context, this paper provides a framework for understanding ``rocket-free maneuvering'' also in a relativistic context. 

Following initial work by Wisdom \cite{Wisdom}, there has been a significant amount of literature already devoted to understanding rocket-free motion in general relativity \cite{Gueron2006, Harte2007, Avron, Gueron2007, Bergamin, Silva2016, PoissonSwim, Vesely2019, VeselyThesis}. The most striking claim which has sometimes been made  is that relativistic extended bodies can ``swim in spacetime'' \cite{Wisdom, Avron}, a description chosen due to similarities with the swimming of microorganisms at low Reynolds numbers \cite{ShapereWilczek, Lauga}: In certain limits, net translations were found to depend only on the sequence of shapes a spacecraft attains, and not on the speed of that sequence. This claim is nevertheless controversial \cite{Silva2016, VeselyThesis}. Although we do not directly address it, understanding motion in a more general context at least provides a basis for investigating whether or not it is possible to swim in spacetime.

Our discussion is initially general but then specializes. Sect. \ref{Sect:Motion} reviews the theory of extended test bodies in general relativity, focusing on the quadrupole approximation and on the simplifications which arise in vacuum (but otherwise arbitrary) spacetimes. Sect. \ref{Sect:TypeD} specializes to vacuum type D geometries, determining all possible quadrupolar forces and torques. General constraints are derived, and the Kerr and Schwarzschild cases are examined in detail. Sect. \ref{Sect:Schw} specializes further, focusing on spin-free, torque-free bodies in the Schwarzschild spacetime. It illustrates how such bodies can use their quadrupole moments to control their motion. There are four appendices. Appendix \ref{App:Conv} explains our notation and conventions and provides a table of symbols. Appendix \ref{App:Kerr} lists coordinate expressions relevant to the Kerr and Schwarzschild geometries. Appendix \ref{App:epDef} reviews the definitions of eccentricity and semi-latus rectum for geodesics in the Schwarzschild spacetime. Finally, Appendix \ref{App:quad} provides some intuition for relativistic quadrupole moments by computing the moment associated with a variable-length rod.

\section{Theory of Motion: A review}
\label{Sect:Motion}

There are many perspectives on motion in general relativity. The one adopted here is due primarily to Dixon \cite{Dixon70a, Dixon74, Dix79, DixonReview, EhlersRudolph}, who derived laws of motion through all multipole orders and without any slow-motion or similar assumptions. Although these laws were originally obtained only in a test body regime, the formalism has since been generalized to allow for nontrivial self-interaction \cite{HarteGrav, HarteReview}.

The discussion here is nevertheless confined  to the consideration of extended test bodies; self-forces and self-torques are ignored. In addition, we assume that (i) a body can be described by a spatially-compact stress-energy tensor $T^{ab}$, and (ii) that $T^{ab}$ must be conserved:
\begin{equation}
	\nabla_b T^{ab} = 0	.
	\label{stressCons}
\end{equation}
Non-gravitational external forces are therefore excluded, as are cases where a body absorbs or ejects material (such as rocket fuel). This section summarizes certain aspects of Dixon's formalism, focusing on the quadrupole approximation and on the simplifications which arise in vacuum backgrounds. Except for some results on mass and momentum quadrupoles in Sect. \ref{Sect:quad} [cf. \eqref{Qremap} and \eqref{Jexpand}], it is a review. More detailed reviews may be found in \cite{Dix79, DixonReview, HarteReview}, although it is the perspective in \cite{HarteReview} which is the closest to the one adopted here.

\subsection{Multipole moments, worldlines, and foliations}

 Fundamentally, Dixon's formalism is a theory of multipole moments. It introduces a set of moments   which are optimized for describing stress-energy tensors, i.e., rank-2, symmetric tensor fields which are also conserved. Had the moments not been constructed with care, stress-energy conservation would imply an evolution equation for each moment. However, it is a central result of Dixon's formalism that with appropriate definitions, the conservation equation constrains  \textit{only} the monopole and dipole moments \cite{Dixon74}. The four partial differential equations \eqref{stressCons} are in fact equivalent to the ten ordinary differential equations which evolve a body's linear and angular momenta. These latter equations are what we refer to as the laws of motion.

As with any multipole expansion, Dixon's constructions require a choice of origin. This takes the form of a worldline $\calZ$, which is assumed here to be timelike. Additionally, there must be a way to identify each point on $\calZ$ with the ``simultaneous'' points within a body's worldtube. Mathematically, this is accomplished by fixing a collection $\Sigma \equiv \{ \Sigma_s | s \} $ of hypersurfaces which foliate the worldtube. Each leaf of that foliation is assumed to intersect $\calZ$ exactly once, so all points in $\Sigma_s$ are identified as being synchronous with the point
\begin{equation}
	z_s \equiv \calZ \cap \Sigma_s
	\label{zsDef}
\end{equation}
on $\calZ$. The time parameter $s$ is arbitrary.

Now, an extended body occupies an extended worldtube. Dixon's constructions effectively replace that worldtube with the worldline $\calZ$. Moreover, they replace $T^{ab}$ with an infinite set of multipole moments. These are tensors on $\calZ$. Different choices for $\calZ$ and $\Sigma$ result in different moments. Although it is possible in principle to transform between different sets of moments, the practical utility of the multipole expansion depends on reasonable choices for $\calZ$ and $\Sigma$. In a Newtonian multipole problem, the analog of $\calZ$ is typically fixed by placing it at a body's center of mass \cite{Dix79, HarteReview}. However, fixing the origin is logically the final step in the theory. Most results do not depend on it, whether in the Newtonian context or in the relativistic one. For this reason, we leave $\calZ$ and $\Sigma$ as arbitrary until describing relativistic center of mass conditions in Sect. \ref{Sect:CM}. Those conditions are applied below in Sect. \ref{Sect:Schw}, but not in Sect. \ref{Sect:TypeD}.

\subsection{Generalized momentum and generalized force}
\label{Sect:genP}

The laws of motion implied by stress-energy conservation affect a body's linear and angular momenta. However, the linear and angular momenta may be viewed as two aspects of a more fundamental object, the \textit{generalized momentum} \cite{HarteSyms} 
\begin{equation}
	\calP_\xi (s) \equiv \int_{\Sigma_s} dS_a T^{a}{}_{b}  \xi^b .  
	\label{PDef}
\end{equation}
This takes as input a vector field $\xi^a$ and a worldline parameter $s$. The allowed vector fields here are known as generalized Killing fields. They are constructed from $\calZ$ and $\Sigma$, and complete definitions may be found in \cite{HarteSyms, Bobbing}. For our purposes, it suffices to note that there is a sense in which the generalized Killing fields preserve geodesic distances away from $\calZ$. They are exactly Killing on $\calZ$, in the sense that
\begin{equation}
	\left. \mathscr{L}_\xi g_{ab} \right|_\calZ = \left. \nabla_c \mathscr{L}_\xi g_{ab} \right|_\calZ = 0.
	\label{Lieg}
\end{equation}
Furthermore, any generalized Killing field is uniquely determined by specifying itself and its (antisymmetric) first derivative anywhere on $\calZ$. Since these data may be specified freely, the space of generalized Killing fields is always ten-dimensional. If an ordinary Killing field exists, it is also a generalized Killing field. These statements together imply that in maximally-symmetric spacetimes, all generalized Killing fields are ordinary Killing fields. 

For fixed $s$, the generalized momentum is a vector not at any tangent space in spacetime, but on the ten-dimensional vector space dual to the space of generalized Killing fields \cite{HarteReview}. The ten components of this vector determine the four components of a body's linear momentum and the six components of its angular momentum. Despite its abstractness, working with the generalized momentum simplifies discussions of conservation laws and allows for calculations which simultaneously describe a body's translational and rotational degrees of freedom. 

The interesting point is now to understand the \textit{generalized force}, which is the rate of change of the generalized momentum. This may be computed using stress-energy conservation: If $w^c$ is a time evolution vector field for $\Sigma$, it follows from \eqref{stressCons} and \eqref{PDef} that
\begin{align}
	\frac{d }{ds} \calP_\xi = \frac{1}{2} \int_{\Sigma_s} T^{ab} \mathscr{L}_\xi g_{ab} w^c dS_c.
	\label{genForceExact}
\end{align} 
This is exact. It illustrates that the generalized force---which determines ordinary forces and torques---measures the degree by which the generalized Killing fields fail to be genuinely Killing. If there exists a genuine Killing field $\psi^a$, it is immediate that the associated component of the generalized momentum is conserved: $\calP_\psi = \mathrm{constant}$.

\subsection{Quadrupole moments}
\label{Sect:quad}

In principle, the generalized force can always be computed using \eqref{genForceExact}. However, directly evaluating  the integral in that expression would require a detailed knowledge of $T^{ab}$. This can be avoided, as much as possible, by assuming (i) that the body is small compared to the lengthscales set by the geometry, and (ii) that $\calZ$ is not too far from the body's ``center.'' With these assumptions, $\Lie_\xi g_{ab}$ can be expanded about $z_s$. The first two terms in that expansion vanish on account of \eqref{Lieg}. The third results in \cite{HarteReview}
\begin{equation}
	\frac{d }{ds} \calP_\xi (s) =  -\tfrac{1}{6} \Jc^{abcd}(s) \mathscr{L}_\xi R_{abcd}(z_s) + \ldots,
	\label{FNvariational}
\end{equation}
where $\Jc^{abcd}$ is the quadrupole moment of $T^{ab}$. This moment depends on both $\calZ$ and $\Sigma$, and has the same algebraic properties as a Riemann tensor:
\begin{equation}
	\Jc^{abcd} = \Jc^{[ab]cd} = \Jc^{ab[cd]} , \qquad \Jc^{[abc]d} = 0 .
	\label{Jsyms}
\end{equation}
All aspects of a body's internal structure which are relevant to its motion are encoded, at this order, in $\calP_\xi$ and in $\Jc^{abcd}$.

Although quadrupole moments are typically used in contexts where a body's small-scale features are not known, they can be computed in terms of those features, when available. The general result \cite{Dixon74} is involved and not reproduced here. However, if Riemann normal coordinates with origin $z_s$ are introduced and denoted by $x^\alpha$, the quadrupole moment can be approximated by
\begin{align}
	\tilde{J}^{\alpha\beta\gamma\delta}(s) = \int_{\Sigma_s} dS_\sigma x^{[\alpha} \big( w^{|\sigma|} T^{\beta][\delta}  + T^{\beta]\sigma} \dot{z}_s^{[\delta} 
	\nonumber
	\\
	~ + \dot{z}^{\beta]}_s T^{\sigma[\delta} \big) x^{\gamma]}  + \ldots  ,
	\label{Jint}
\end{align}
where the omitted terms here have relative magnitude $[\mbox{(body size)}/\mbox{(curvature scale)}]^2$. Appendix \ref{App:quad} applies this to compute the quadrupole moment of a rod whose length is a freely-specifiable function of time.

In general, \eqref{Jsyms} implies that $\tilde{J}^{abcd}$ has twenty independent components.  However, not all of those components affect the motion. In vacuum geometries, at least ten of them are irrelevant \cite{HarteTrenorm, HarteGrav, Bini2014}. This may be seen by noting that if $R_{ab} = R_{acb}{}^{c} = 0$, and if
\begin{align}
	J^{abcd} \equiv (\Jc^{abcd})_\mathrm{TF}
	 = \Jc^{abcd} + g^{b[c} \Jc^{d]fa}{}_{f}
	 	\nonumber
	\\ 
	~ - g^{a[c} \Jc^{d]fb}{}_{f}  + \tfrac{1}{3} \Jc^{fh}{}_{fh} g^{a[c} g^{d]b}
	\label{JTF}
\end{align}
denotes the trace-free component of $\Jc^{abcd}$, it follows from \eqref{Lieg} that\footnote{Eq. \eqref{TF} is valid also for vacuum solutions with nonzero cosmological constant $\Lambda$, in which case $R_{ab} = \Lambda g_{ab}$. Nevertheless, we assume $\Lambda = 0$ below.}
\begin{equation}
	(J^{abcd} -  \Jc^{abcd} ) \Lie_\xi R_{abcd} = 0.	
	\label{TF}
\end{equation}
Comparison with \eqref{FNvariational} then shows that the trace-free moment $J^{abcd}$ can be used to compute the same forces and torques as the full moment $\Jc^{abcd}$; the ten trace components $\Jc^{acb}{}_{c}$ are irrelevant. However, even this does not complete the counting of irrelevant quadrupole components. It is shown in Sect. \ref{Sect:TypeD} below that in vacuum spacetimes which are of Petrov type D, at least four additional components decouple from the laws of motion. 

As alluded to above, stress-energy conservation imposes no differential constraints on the evolution of $\tilde{J}^{abcd}$ \cite{Dixon74}. It follows that there are also no universal evolution equations for the trace-free moment $J^{abcd}$. Except where forbidden by energy conditions, there exist conserved stress-energy tensors---and therefore physically-realizable bodies---in which $J^{abcd}$ evolves in any manner whatsoever (so long as it retains the algebraic symmetries of a Weyl tensor). This is essential for the discussion in Sect. \ref{Sect:Schw} below, as it ensures that at least in principle, spacecraft can be engineered to have quadrupole moments which vary in any specified manner. In particular, they may be engineered to control their quadrupole moments in order to control their motion.

Some intuition for $J^{abcd}$ may be gained by decomposing it into its mass and momentum components. Any such decomposition first requires a choice of frame. Choosing one by fixing a unit timelike vector $\tau^a$, existing decompositions for $\Jc^{abcd}$, which originated in \cite{EhlersRudolph}, have employed a mass quadrupole $\tilde{Q}^{ab} = \tilde{Q}^{(ab)}$, a rank-3 momentum quadrupole $\tilde{\Pi}^{abc} = \tilde{\Pi}^{a[bc]}$, and a rank-4 stress quadrupole $\tilde{S}^{abcd} = \tilde{S}^{[ab]cd} = \tilde{S}^{ab[cd]}$. These tensors satisfy $\tilde{\Pi}^{[abc]} = \tilde{S}^{[abc]d} = 0$ and are fully orthogonal to $\tau^a$. In terms of them,
\begin{equation}
	\Jc^{abcd} = \tilde{S}^{abcd} - \tau^{[a} \tilde{\Pi}^{b]cd} - \tau^{[c} \tilde{\Pi}^{d]ab} - 3 \tau^{[a} \tilde{Q}^{b][c} \tau^{d]} .
	\label{JexpandFull}
\end{equation}
In non-vacuum backgrounds, the relative complexity of this decomposition is essential. In the vacuum context of interest here, it can be simplified. This is because \eqref{TF} allows $\tilde{J}^{abcd}$ to be replaced by $J^{abcd}$ in the laws of motion, and for the latter, the stress quadrupole can be discarded, the trace of the mass quadrupole is irrelevant, and the rank-3 momentum quadrupole can be replaced by a rank-2 momentum quadrupole. More precisely, it is useful to define a new set of mass and momentum quadrupoles by
\begin{subequations}
\label{Qremap}
\begin{align}
	Q^{ab} &\equiv (\tilde{Q}^{ab} + \tfrac{4}{3} \tilde{S}^{acb}{}_{c} )_\mathrm{TF}  , 
	\\
	\Pi^{ab} &\equiv -\tilde{\Pi}^{(a}{}_{cd} \epsilon^{b)cdf} \tau_f,
\end{align}
\end{subequations}
where the trace-free operation in the first line is to be performed using the spatial projector $g^{ab} + \tau^a \tau^b$. Both $Q^{ab}$ and $\Pi^{ab}$ are symmetric, trace-free, and orthogonal to $\tau^a$. Each has five independent components. Use  of \eqref{JTF} and \eqref{JexpandFull} shows that they completely determine the trace-free stress-energy quadrupole:
\begin{align}
	J^{abcd} = \big[ \tfrac{1}{2} ( \tau^{[a} \Pi^{b]f} \epsilon^{cd}{}_{fh} + \tau^{[c} \Pi^{d]f} \epsilon^{ab}{}_{fh}  ) \tau^h 
	\nonumber
	\\
	~ - 3 \tau^{[a} Q^{b][c} \tau^{d]} \big]_\mathrm{TF}.
		\label{Jexpand}
\end{align}
Note that this is an expansion for $J^{abcd}$ while   \eqref{JexpandFull} is an expansion for $\tilde{J}^{abcd}$.

\subsection{Linear momentum, angular momentum, force, and torque}

We now use the generalized momentum $\calP_\xi$ to define a body's linear momentum $p_a$ and its angular momentum $S^{ab} = S^{[ab]}$, both of which are tensors on $\calZ$. The definition is implicit. For all generalized Killing fields \cite{HarteSyms, HarteReview},
\begin{equation}
	\calP_\xi(s) = p_a(s) \xi^a(z_s) + \tfrac{1}{2} S^{ab}(s) \nabla_a \xi_b(z_s) .
	\label{pSDef}
\end{equation}
For fixed $\xi^a$, this shows that $\calP_\xi$ is a linear combination of linear and angular momentum components. All components of $p_a$ and $S^{ab}$ may be extracted by varying over the full ten-dimensional space of generalized Killing fields. Alternatively, explicit integrals for the momenta may be found in \cite{Dixon70a, Dixon74, HarteReview}.

A force $F_a$ and a torque $N^{ab} = N^{[ab]}$ can be defined similarly, using the generalized force $d \calP_\xi/ds$. For all generalized Killing fields,
\begin{equation}
	\frac{d}{ds} \calP_\xi(s) = F_a(s) \xi^a(z_s) + \tfrac{1}{2} N^{ab}(s) \nabla_a \xi_b(z_s) .
	\label{FNDef}
\end{equation}
Evolution equations for $p_a$ and $S^{ab}$ follow by differentiating \eqref{pSDef}, comparing the result with \eqref{FNDef}, and then varying over all generalized Killing fields. Using $(\nabla_c \nabla_a \xi_b + R_{abcd} \xi^d) \big|_\calZ = 0$, which follows from \eqref{Lieg}, these steps result in Dixon's equations \cite{Dixon74, Dix79, HarteSyms, HarteReview}
\begin{subequations}
\label{Dixon}
\begin{align}
	\dot{p}_a  &= -\tfrac{1}{2} R_{abcd} \dot{z}_s^b S^{cd} + F_a, 
	\label{pDot}
	\\
	\dot{S}^{ab} &= 2 p^{[a} \dot{z}_s^{b]} + N^{ab}.
	\label{Sdot}
\end{align}
\end{subequations}
Furthermore, comparison of \eqref{FNvariational}, \eqref{TF}, and \eqref{FNDef} shows that in the quadrupole approximation, the force and torque are given by
\begin{align}
	F_a = - \tfrac{1}{6} J^{bcdf} \nabla_a R_{bcdf} , \quad N^{ab} = \tfrac{4}{3} J^{cdf[a} R^{b]}{}_{fcd} 
	\label{FN}
\end{align}
in vacuum spacetimes. Contributions from the octupole and higher-order moments may be found in \cite{Dixon74, Dix79, HarteReview} (without restriction to the vacuum case).

As already noted, Killing fields generate conservation laws; if $\psi^a$ is Killing,  $\calP_\psi$ is conserved. Using \eqref{pSDef}, this is equivalent to the conservation of a linear combination of linear and angular momentum components:
\begin{equation}
	\calP_\psi = p_a \psi^a + \tfrac{1}{2} S^{ab} \nabla_a \psi_b = \mathrm{constant}. 
	\label{genCons}
\end{equation}
It also follows from \eqref{FNDef} that for each Killing field, a linear combination of force and torque components must vanish. In particular, 
\begin{equation}
	F_a \psi^a + \tfrac{1}{2} N^{ab} \nabla_a \psi_b =0.
		\label{fnCons}
\end{equation} 
These results are exact. They are also preserved at every multipole order \cite{EhlersRudolph}.

It may be noted that our force and torque are not necessarily equal to $\dot{p}_a$ and $\dot{S}^{ab}$. Although the difference terms in \eqref{Dixon} are sometimes referred to as forces and torques, this is physically inappropriate \cite{Dix79, HarteReview}. They are purely kinematic consequences of the fact that a local Poincar\'{e} transformation at $z_s$ may look different from that same transformation at $z_{s+ds}$. For example, a local Lorentz transformation about one point is equivalent to a Lorentz transformation about another point \textit{together with} a translation. Similarly, a local translation at one point is equivalent, at another point, to a local translation \textit{together with} a local Lorentz transformation. These geometric effects make the linear and angular momenta appear to mix when evaluated at different points on $\calZ$; they are responsible for the $-\frac{1}{2} R_{abcd} \dot{z}^b_s S^{cd}$ and $2p^{[a} \dot{z}^{b]}_s$ terms in Dixon's equations. Such terms  do not affect the generalized momentum, whose variations are intrinsically dynamical. Indeed, it is only these dynamical variations which we refer to as forces and torques.

\subsection{Center of mass and the momentum-velocity relation}
\label{Sect:CM}

Everything said so far has been valid regardless of the  choice of worldline $\calZ$ or the foliation $\Sigma$. In fact, the foliation plays no explicit role in the application of the formalism. It arises only when relating the momenta or the moments to the stress-energy tensor.  Nevertheless, $\Sigma$ can be fixed by letting each $\Sigma_s$ be formed from the set of spacelike geodesics which emanate from $z_s$ and are orthogonal at that point to $p^a(s)$. Furthermore, $\calZ$ may be fixed by supposing that \cite{EhlersRudolph, Dix79}
\begin{equation}
	S^{ab} p_b = 0.
	\label{SSC}
\end{equation}
This can be interpreted as requiring that the mass dipole moment vanish for a zero-momentum observer. As both $p_a$ and $S^{ab}$  depend on $\calZ$ and $\Sigma$, these definitions are highly implicit. Nevertheless, it has been shown that under appropriate conditions, solutions exist and are unique \cite{CM1, CM2}. We refer to the resulting $\calZ$ as the center of mass worldline.

The center of mass velocity $\dot{z}^a_s$ may now be related to the momentum $p^a$. In Newtonian physics, the analogs of these quantities are proportional to one another. However, that is not a definition. It is instead a consequence of the definitions for the momentum and the center of mass position. Relativistically, the relation between momentum and center of mass velocity must also be derived. This was first accomplished by Ehlers and Rudolph \cite{EhlersRudolph}, who showed that these quantities are not necessarily proportional. To describe their result, first write the momentum in terms of a (not necessarily constant) rest mass $\calM \equiv \sqrt{ - p_a p^a} > 0$ and a unit vector $u^a$, such that
\begin{equation}
	p_a = \calM u_a	, \qquad 	u_a u^a = -1.
	\label{mDef}
\end{equation}
If the worldline parameter $s$ is then normalized such that
\begin{equation}
	\dot{z}^a_s u_a = -1		,
	\label{sDef}
\end{equation}
\eqref{Dixon} and \eqref{SSC} may be shown to imply that 
\begin{align}
	\calM \dot{z}^a_s &= p^a - N^{a}{}_{b} u^b 
	\nonumber
	\\
	& ~ - \frac{ S^{ab} [ \calM F_b - \tfrac{1}{2} ( p^c - N^{c}{}_{e} u^e) S^{df} R_{bcdf} ] }{ \calM^2 + \tfrac{1}{4} S^{bc} S^{df} R_{bcdf}}.
	\label{mvRelation}
\end{align}
Note that $\dot{z}^a_s$ does not appear on the right-hand side of this expression. In general, $s$ is not a proper time and $u^a$ is not tangent to $\dot{z}^a_s$. The difference $\calM \dot{z}_s^a - p^a$ is nevertheless orthogonal to $p^a$, and is referred to as the hidden mechanical momentum \cite{Bobbing, CostaReview}. Also, it is assumed here that the denominator $\calM^2 + \tfrac{1}{4} S^{bc} S^{df} R_{bcdf}$ never vanishes in any situation where this equation is to be applied. If it did vanish, that would signal a breakdown of the center of mass condition.

When the center of mass condition is applied, $S^{ab}$ can have only three nonzero components. Nevertheless, the torque $N^{ab}$ is not similarly constrained. Three of its six components directly contribute to changes in the angular momentum. Inspection of \eqref{mvRelation} shows that the remaining three torque components, determined by $N^{a}{}_{b}u^b$, contribute to the hidden momentum.

\section{Extended-body effects in type D spacetimes}
\label{Sect:TypeD}

Although our interest is primarily in the Kerr and Schwarzschild geometries, many results can be derived just as easily while assuming only that the spacetime is vacuum and of Petrov type D. This includes Kerr as a special case, although it also allows for accelerating black holes, objects with nonzero NUT charge, and more \cite{Plebanski, GriffithsExact}. This section examines the gravitational forces and torques which act on extended bodies in arbitrary vacuum type D spacetimes. It does not impose any center of mass conditions. 

\subsection{Geometry of type D spacetimes}

Simple expressions for forces and torques require decompositions adapted to the spacetime geometry. More precisely, they require a tetrad adapted to the principal null directions of that geometry. By definition, there are two such directions in type D spacetimes. Choosing the real null vectors $\ell^a$ and $n^a$ to be tangent to those directions, it is convenient to introduce a complex null vector $m^a$ such that $(\ell^a, n^a, m^a, \bar{m}^a)$ is a  tetrad whose only non-vanishing scalar products are 
\begin{equation}
	m \cdot \bar{m} = - \ell \cdot n = 1.
	\label{tetradProd}
\end{equation}
This implies that $g_{ab} = 2 [m_{(a} \bar{m}_{b)} - \ell_{(a} n_{b)}]$. Tetrads with these properties are unique up to the discrete swaps $\ell^a \leftrightarrow n^a$ and $m^a \leftrightarrow \bar{m}^a$, and the rescalings
\begin{equation}
	\ell^a \mapsto \lambda \ell^a, \qquad n^a \mapsto \lambda^{-1} n^a, \qquad m^a \mapsto e^{i \zeta} m^a,
	\label{rescale}
\end{equation}
where $\lambda \neq 0$ and $\zeta$ are real but otherwise arbitrary.

Fixing any tetrad in this class, it is convenient to define from it a basis of complex 2-forms, given by
\begin{equation}
\begin{gathered}
	X_{ab} = 2 \ell_{[a} m_{b]} , \qquad Y_{ab} = 2n_{[a} \bar{m}_{b]} ,
	\\
	Z_{ab} = 2 ( \ell_{[a} n_{b]} - m_{[a} \bar{m}_{b]} )  
\end{gathered}
\label{2formBasis}
\end{equation}
and their complex conjugates. The only non-vanishing inner products in this basis follow from
\begin{equation}
	Z_{ab} Z^{ab} = 2 X_{ab} Y^{ab} = -4.
	\label{bivectProd}
\end{equation}
It may be noted that $i Z_{ab}$ is a square root of the metric in the sense that $g_{ab} = - Z_{a}{}^{c} Z_{bc}$. Additionally, the basis elements $X_{ab}$, $Y_{ab}$, and $Z_{ab}$ are self-dual, meaning that, e.g., $X^*_{ab} = i X_{ab}$, where $X^*_{ab} \equiv \frac{1}{2} \epsilon_{ab}{}^{cd} X_{cd}$ denotes the Hodge dual. The conjugate basis elements $\bar{X}_{ab}$, $\bar{Y}_{ab}$, and $\bar{Z}_{ab}$ are anti self-dual, so, e.g., $\bar{X}_{ab}^* = - i \bar{X}_{ab}$. Our main motivation for introducing this basis is that it allows the curvature (and later the quadrupole moment) to be written down and manipulated without having to perform  coordinate computations.

In order to write down the curvature, first note that since $\ell^a$ and $n^a$ are both tangent to repeated principal null directions, 
\begin{equation}
	\ell_{[a} R_{b]cdf} \ell^c \ell^d = n_{[a} R_{b]cdf} n^c n^d = 0.
	\label{PNDs}
\end{equation}
These equations and the vacuum condition $R_{ab} = 0$ can be used to show that the Riemann tensor is fixed up to a complex scalar $\PSi$:
\begin{equation}
	R_{abcd} = 2 \Re \big[ \PSi (Z_{ab} Z_{cd} - X_{ab} Y_{cd} - Y_{ab} X_{cd} ) \big].
	\label{Rcurv}
\end{equation}
Here, $\PSi = - \frac{1}{4} R_{abcd} X^{ab} Y^{ab}$ is more commonly denoted by $\Psi_2$, and is one of five Weyl scalars $\Psi_0, \ldots, \Psi_4$ \cite{Hall, Sachs}. In type D spacetimes and with a tetrad of the given type, the other scalars vanish and are not used below. 

Many of the most mathematically-interesting characteristics of type D spacetimes follow from the fact that they admit a Killing spinor $\kappa_{AB} = \kappa_{(AB)}$, which is defined to satisfy $\nabla_{A'(A} \kappa_{BC)} = 0$ \cite{LarsSpinReview}. The relevant point for our purposes is that the existence of this spinor implies the existence of two real conformal  Killing-Yano tensors. These are the real and imaginary components of the complex 2-form corresponding to $\kappa_{AB} \bar{\epsilon}_{A'B'}$. In order to write this down more explicitly, it is convenient to introduce a spinor dyad $(o^A, \iota^A)$ such that $\ell^a = o^A \bar{o}^{A'}$ and $n^a = \iota^A \bar{\iota}^{A'}$, in which case $\kappa_{AB} = \PSi^{-1/3} o_{(A} \iota_{B)}$  \cite{WalkerPenrose}. Introducing an arbitrary constant $\chi$ for later convenience, a complex conformal Killing-Yano tensor derived from $\kappa_{AB}$ is then
\begin{equation}
	\CKY_{ab} \equiv \chi \PSi^{-1/3} Z_{ab}	 .
	\label{CKY}
\end{equation}
This satisfies the conformal Killing-Yano equation
\begin{equation}
	\nabla_{(a} \CKY_{b)c} = g_{ab} \CKY_c - \CKY_{(a} g_{b)c},
\end{equation}
where $\CKY_a \equiv \frac{1}{3} \nabla^b \CKY_{ba}$. As $Z_{ab}$ is self-dual, so is $\CKY_{ab}$. The square of a conformal Killing-Yano tensor is a rank-2 conformal Killing tensor, which generates quadratic conservation laws for null geodesics. Although there do not appear to be generalizations of these conservation laws which apply for massive extended bodies, $\CKY_{ab}$ will nevertheless be seen to play an important role in the analysis of their motion.

Before explaining this, it is instructive to provide an example of the structures just described. The Kerr spacetime with mass $M$ and specific angular momentum $a$ is vacuum and type D. Its metric is given, in Boyer-Lindquist coordinates $(t,r,\theta ,\phi)$, by \eqref{Kerr} below. An explicit null tetrad is provided by \eqref{tetradKerr}, and in terms of that, the Weyl scalar appearing in \eqref{Rcurv} is
\begin{align}
	\PSi  = - \frac{M}{(r - i a \cos \theta)^3} .
	\label{Psi2}
\end{align}
In the $a=0$ Schwarzschild case, $\Psi$ is real. Otherwise, it is complex. Turning to the conformal Killing-Yano tensor $\CKY_{ab}$ and allowing for arbitrary $a$, it is convenient to choose the $\chi$ in \eqref{CKY} such that
\begin{equation}
	\CKY_{ab} = i (r- i a \cos \theta) Z_{ab}.
	\label{CKYKerr}
\end{equation}
In coordinates, this is given by \eqref{CKYcoords}. Its real component is an ordinary (divergence-free) Killing-Yano tensor, the square of which is the rank-2 Killing tensor which determines the Carter constants for Kerr geodesics.

\subsection{Quadrupolar forces and torques in type D spacetimes}
\label{Sect:FN}

We now compute forces and torques in general vacuum type D spacetimes, allowing for arbitrary quadrupole moments but no octupole or higher moments. Contrary to common practice, we do not decompose the quadrupole moment into mass and momentum components. Instead, we observe that simple results for the force and torque arise when the quadrupole moment is expressed in a form which is adapted to the background geometry, and not to, e.g., a body's rest frame.

The decomposition adopted here takes advantage of the fact that the trace-free quadrupole moment $J_{abcd}$ has the same algebraic properties as a Weyl tensor. It is therefore possible to decompose it into the five complex scalars
\begin{subequations}
\label{Ji}
\begin{gather}
	J_0 \equiv \tfrac{1}{4} J_{abcd} X^{ab} X^{cd}, \quad J_1 \equiv \tfrac{1}{8} J_{abcd} X^{ab} Z^{cd},
	\\
	J_2 \equiv -\tfrac{1}{4} J_{abcd} X^{ab} Y^{cd} = \tfrac{1}{16} J_{abcd} Z^{ab} Z^{cd},
	\\
	J_3 \equiv -\tfrac{1}{8} J_{abcd} Y^{ab} Z^{cd}, \quad J_4 \equiv \tfrac{1}{4} J_{abcd} Y^{ab} Y^{cd},
\end{gather}
\end{subequations}
which are analogous to the five Weyl scalars $\Psi_0, \ldots, \Psi_4$. These definitions are equivalent to the expansion
\begin{align}
	J_{abcd} = 2 \Re \big[ J_0 Y_{ab} Y_{cd} + J_1 (Y_{ab} Z_{cd} + Z_{ab} Y_{cd}) 
	\nonumber
	\\
	~ + J_2 (Z_{ab} Z_{cd} - X_{ab} Y_{cd} - Y_{ab} X_{cd} )
	\nonumber
	\\
	~ -  J_3 ( X_{ab} Z_{cd} + Z_{ab} X_{cd}) + J_4 X_{ab} X_{cd} \big].
	\label{quad}
\end{align}

It is now straightforward to compute quadrupolar forces and torques for arbitrary extended bodies: Combining \eqref{Lieg}, \eqref{FNvariational}, \eqref{bivectProd}, \eqref{Rcurv}, \eqref{CKY}, and \eqref{Ji} shows that the generalized force dual to any generalized Killing field $\xi^a$ is 
\begin{align}
	\frac{ d  }{ ds } \calP_\xi &= F_a \xi^a + \tfrac{1}{2} N^{ab} \nabla_a \xi_b 
	\nonumber
	\\
	&= -8 \Re  \big[ J_2 \mathscr{L}_\xi \PSi   
	+ \tfrac{1}{2} \chi^{-1} \PSi^{4/3} ( J_3 X^{ab} 
	\nonumber
	\\
	& \qquad \qquad \qquad \qquad \qquad ~ - J_1 Y^{ab}  )  \mathscr{L}_\xi \CKY_{ab}   \big].
	\label{genForce}
\end{align}
This is valid for bodies with arbitrary internal structure and for any $\Sigma$ and $\calZ$. The only assumption is that the generalized force can be truncated at quadrupole order. Regardless, the result splits into two parts: one  proportional to $\Lie_\xi \PSi$ and the other to $\Lie_\xi \CKY_{ab}$. These terms measure the degrees by which $\xi^a$ fails to generate symmetries for the Weyl scalar $\PSi$ or the conformal Killing-Yano tensor $\CKY_{ab}$.

One consequence of \eqref{genForce} is that since $J_0$ and $J_4$ are absent from that expression, they cannot affect a body's motion. It is only $J_1$, $J_2$, and $J_3$ which influence $d \calP_\xi/ds$. The ten real force and torque components are thus determined by three complex quadrupole components. This means that there must be a minimum\footnote{If $\PSi$ is real, as it is in Schwarzschild, $\Im J_2$ cannot affect the motion. There are then five force and torque constraints instead of four.} of four real constraints on the force and torque. Type D spacetimes admit either two or four Killing fields \cite{Kinnersley}, so at least in cases with only two Killing fields, these constraints cannot only be of the form \eqref{fnCons}. There must be additional constraints which are not derivable from Killing fields.

To identify these additional constraints, it can be useful to work with $F_a$ and $N^{ab}$ instead of $d \calP_\xi/ds$. Varying \eqref{genForce} over all generalized Killing fields while noting that $X^{ab} \mathscr{L}_\xi Z_{ab} = 4 m_a \mathscr{L}_\xi \ell^a$ and $Y^{ab} \mathscr{L}_\xi Z_{ab} = -4 \bar{m}_a \mathscr{L}_\xi n^a$, 
\begin{align}
	F_a = -8  \Re \big[ J_2 \nabla_a \PSi  +2 \PSi ( J_1 \bar{m}^b \nabla_a n_b  
	 \nonumber
	 \\
	 ~ + J_3 m^b \nabla_a \ell_b )\big],
	 \label{quadF}
\end{align}
and
\begin{equation}
	N^{ab} = 16  \Re \left[ \PSi( J_1 Y^{ab} + J_3 X^{ab})\right] .
	\label{quadN}
\end{equation}
Thus, while the force depends on $J_1$, $J_2$, and $J_3$, the torque depends only on $J_1$ and $J_3$.  Additionally, combining \eqref{quadN} with \eqref{bivectProd} and \eqref{CKY} shows that
\begin{equation}
	 N^{ab} \CKY_{ab} = 0.
	 \label{NW}
\end{equation}
At least when the spacetime admits only two Killing fields, \textit{this is the constraint which is not derivable from Killing vectors}. It implies that two real torque components must vanish, regardless of a body's internal structure.

Another perspective on force and torque constraints may be gained by noting that
\begin{align}
	F_a = \Re \big[ \bar{m}^b  \big(  X_{cd} \nabla_a n_b +
	  \bar{Y}_{cd} \nabla_a \ell_b  \big) \big] N^{cd}  
	  \nonumber
	  \\
	  ~-8\Re (J_2 \nabla_a \PSi ),
	  \label{forcetorque}
\end{align}
which follows from \eqref{quadF} and \eqref{quadN}. If this equation is contracted with a Killing field, \eqref{PNDs} and \eqref{NW} can be used to show that it implies the Killing constraint \eqref{fnCons}. More generally, this can be viewed as describing that component of the force which may be varied independently of the torque: If a spacecraft has actively adjusted $J_1$ and $J_3$ to produce a desired torque---subject to the constraint \eqref{NW}---the force can be varied only via $\Re (J_2 \nabla_a \PSi)$. As $\mathscr{L}_\psi \PSi = 0$ for any Killing vector $\psi^a$, forces can thus be controlled, independently of torques, only in directions which are ``not Killing.'' This provides an intuitive sense in which a body can ``grab on to the geometry'' only in directions where the geometry is changing. 

\subsection{Conservation laws in Kerr}

We now specialize to a Kerr spacetime with mass $M$ and specific angular momentum $a$. As given by \eqref{KillingKerr}, there are two Killing vectors  in these spacetimes: $t^a$, which generates a time translation,  and $\psi^a_{(3)}$, which generates a rotation. Each of these Killing vectors implies a conservation law and also a constraint on the force and torque. 

Beginning with $t^a$, it follows from \eqref{genCons}, \eqref{Psi2}, and \eqref{CKYKerr} that the energy
\begin{equation}
	E \equiv -  \calP_t = - p_a t^a + \tfrac{1}{2} \Im \left( \PSi S^{ab} \CKY_{ab} \right) 
	\label{energy}
\end{equation}
must be conserved, regardless of a body's internal dynamics. Using \eqref{fnCons}, the corresponding constraint on the force and torque is 
\begin{equation}
	F_a t^a = \tfrac{1}{2} \Im \left( \PSi N^{ab} \CKY_{ab} \right) .
	\label{FtCons}
\end{equation}
Both this and \eqref{energy} hold through all multipole orders. However, use of \eqref{NW} shows that in the quadrupole approximation,
\begin{equation}
	F_a t^a = 0.
	\label{FtZero}
\end{equation}

Similar calculations may be performed using the rotational Killing field $\psi^a_{(3)}$. Its existence implies that the angular momentum component $\calP_{\psi_{(3)}}$ must be conserved, and also that $F_a \psi^a_{(3)} + \tfrac{1}{2} N_{ab} \nabla^a \psi^b_{(3)} = 0$. In the quadrupole approximation where \eqref{NW} holds, the force and torque constraint is more explicitly
\begin{align}
	F_a \psi^a_{(3)} = \tfrac{1}{2} \Im \left[  \left( \frac{r^2-2M r+a^2}{(r-i a \cos \theta)^2 } \right) X_{ab} + Y_{ab}\right] 
	\nonumber
	\\
	~ \times N^{ab} \sin\theta.
\end{align}

\subsection{Conservation laws in Schwarzschild}
\label{Sect:syms}

We now discuss forces and torques in a Schwarzschild spacetime with mass $M$. In standard Schwarzschild coordinates $(t,r,\theta,\phi)$, the metric components are given by the $a=0$ case of the general Kerr expression \eqref{Kerr}. Unlike other members of the Kerr family, Schwarzschild spacetimes admit four Killing vector fields: $t^a$, which generates time translations, and $\psi^a_{(1)}$, $\psi^a_{(2)}$, and $\psi^a_{(3)}$, which all generate rotations. The coordinate components of these vector fields are given by \eqref{Killing}.

It is convenient to employ a 3-vector notation where the rotational Killing fields are viewed as elements of the triple
\begin{equation}
	\vec{ \psi }^a \equiv \big( \psi_{(1)}^a, \psi_{(2)}^a, \psi_{(3)}^a \big).
\end{equation}
The existence of these Killing fields then implies that the angular momentum ``3-vector''
\begin{equation}
	\vec{L} \equiv \calP_{ \vec{\psi} } \equiv (\calP_{\psi_{(1)} },\calP_{\psi_{(2)} }, \calP_{\psi_{(3)} })
	\label{consLaws}
\end{equation}
must be conserved. To better understand the implications of this conservation law, it is useful to introduce a number of additional definitions. First, motivated by standard transformations between Cartesian and polar coordinates in $\mathbb{R}^3$, define the 3-vector basis
\begin{subequations}
\label{thetaphiDef}
\begin{align}
	\vec{z} &\equiv r \left( \sin \theta \cos \phi, \sin \theta \sin \phi , \cos \theta \right) ,
	\\
	\vec{\theta} &\equiv \left( \cos \theta \cos \phi, \cos \theta \sin \phi, -\sin \theta\right),
	\\
	\vec{\phi} &\equiv \left( - \sin \phi, \cos \phi, 0 \right).
\end{align}
\end{subequations}
If a linear momentum 3-vector is then introduced via\footnote{Any terms in $\vec{p}$ which may be proportional to $\vec{z}$  are irrelevant for our purposes and are excluded.}
\begin{equation}
	\vec{p} \equiv \sqrt{2}  \Re \big[ p_a m^a ( \vec{\theta} - i \vec{\phi} ) \big] ,
	\label{vecP}
\end{equation}
and if ``$\times$'' is used to denote the standard cross product on  $\mathbb{R}^3$, a calculation shows that
\begin{equation}
	p_a \vec{\psi}^a = \vec{z} \times \vec{p} .
\end{equation}
Combining  this with \eqref{genCons} and \eqref{consLaws} finally shows that by defining
\begin{equation}
\vec{S} \equiv \tfrac{1}{2} S^{ab} \nabla_a \vec{\psi}_b,
	\label{sphSym}
\end{equation}
we recover the \textit{Newtonian relation} 
\begin{align}
	\vec{L} = \vec{z} \times \vec{p} + \vec{S}
	\label{Jcons}
\end{align}
between different types of angular momenta. Given this, it is natural to interpret $\vec{L}$, $\vec{z} \times \vec{p}$, and $\vec{S}$ as the total, orbital, and spin angular momenta, respectively. Only $\vec{L}$ is necessarily conserved.

Nearly identical calculations can be used to describe constraints on the forces and torques which are imposed by the rotational Killing fields: If force and torque 3-vectors are defined such that
\begin{equation}
	\vec{z} \times \vec{F} \equiv F_a \vec{\psi}^a, \qquad 	\vec{N} \equiv \tfrac{1}{2} N^{ab} \nabla_a \vec{\psi}_b,
\end{equation}
it follows from \eqref{fnCons} that
\begin{equation}
	\vec{N} + \vec{z} \times \vec{F} = 0.
	\label{FNpsi}
\end{equation}
This and \eqref{Jcons} are valid through all multipole orders.

Additional insight may be gained by examining the components of these equations which lie  parallel to $\vec{z}$. For this purpose, it is first useful to note that
\begin{align}
	\nabla_a \vec{\psi}_b \equiv \Re 
	\big[ \left( (1-2M/r) X_{ab} - \bar{Y}_{ab} \right) (  \vec{\phi} + i \vec{\theta} ) 
	\nonumber
	\\
	~ - \CKY_{ab} (\vec{z}/r^2) \big] .
	\label{scrS}
\end{align}
Then, dotting \eqref{Jcons} with $\vec{z}$ while using \eqref{sphSym} shows that
\begin{equation}
	 (\vec{z}/r) \cdot \vec{L} = - \frac{1 }{2 r} \Re (  S^{ab} \CKY_{ab} ) .
\end{equation}
The left-hand side here is proportional to the cosine of the angle between $\vec{z}$ and $\vec{L}$. Changes in this angle therefore require changes in $\Re (  S^{ab} \CKY_{ab} )/r$. If there is no spin, $\vec{z}$ must lie in the plane orthogonal to $\vec{L}$.

A similar calculation applied to the force and torque constraint \eqref{FNpsi} shows that $\vec{z} \cdot \vec{N} = 0$. Equivalently,
\begin{equation}
	\Re ( N^{ab} \CKY_{ab}) = N^{ab} m_a \bar{m}_b = N^{\theta \phi} = 0.
	\label{NW2}
\end{equation}
This might appear to be a weaker form of the torque constraint \eqref{NW}. However, the two results have different regimes of validity. Eq. \eqref{NW2} was obtained using the three rotational Killing fields in Schwarzschild, and is valid through all multipole orders. By contrast, the derivation of \eqref{NW} assumed less about the spacetime---requiring only that it be vacuum and type D---but more about the multipole structure---requiring that all contributions beyond the quadrupole be ignorable. The conclusion here is that in Schwarzschild, the real component of \eqref{NW} is in fact exact, and may be viewed as a consequence of the Killing constraints. The imaginary component of that equation, $\Im ( N^{ab} \CKY_{ab}) = N^{tr} = 0$, is independent of the Killing constraints and may be violated beyond quadrupole order. 

In Schwarzschild, there are  five force and torque constraints in the quadrupole approximation. Four of these constraints are due to the Killing fields and one to the imaginary  component of \eqref{NW}. In more general Kerr spacetimes with $a \neq 0$, there are instead four force and torque constraints in the quadrupole approximation. Two of these are due to the Killing fields and two to the real and imaginary components of \eqref{NW}. This distinction between the Kerr and Schwarzschild cases is summarized in Table \ref{Table:nums}.

\begin{table}
\setlength{\tabcolsep}{8pt}
  \begin{tabular}{l | c  c  c }
  \hline\hline
   Spacetime & Exact & Quadrupole & Total \\
    \hline
    Kerr 			&	2 &	2 &	4 \\
    Schwarzschild 	& 	4 &	1 &	5\\
    \hline\hline
  \end{tabular}
  \caption{Numbers of real force and torque constraints in Kerr and Schwarzschild spacetimes. Exact constraints are of the form \eqref{fnCons} and follow from Killing fields. Constraints which are necessarily valid only at quadrupole order are of the form \eqref{NW}. Differing numbers of quadrupole constraints are due to the fact that in Schwarzschild, the real component of \eqref{NW} is not independent from the Killing constraints.}
  \label{Table:nums}
\end{table}

\section{Actively-controlled motion in the Schwarzschild spacetime}
\label{Sect:Schw}

Our focus in Sect. \ref{Sect:TypeD} was on determining which forces and torques could (or could not) be produced by appropriately-structured bodies. This amounted to finding constraints on the possible equations of motion. However, except with the conservation laws implied by Killing fields, it is not obvious how constraints on the equations of motion translate into constraints on their solutions. This section discusses some of those solutions.

Doing so requires specialization: We restrict to spin-free, torque-free bodies in Schwarzschild. These bodies are assumed to actively control their quadrupole moments, both to maintain a torque-free state and to control their motion. If octupole and higher moments are ignored, we shall see that the orbits of such bodies are controlled only by $\Re J_2$. That component of the quadrupole moment can be varied to control a body's mass, and from that, radial falls can be slowed or accelerated, unstable orbits can be stabilized, and bound orbits can change their eccentricities.

The analysis here may be viewed as the relativistic generalization of the Newtonian discussion in \cite{HarteNewtonian}. In that context, bodies were also assumed to be torque-free, in part to avoid maneuvers in which a spacecraft would be likely to spin itself apart. The same argument could be applied also in the relativistic context to motivate the torque-free condition. However, setting $N^{ab} = 0$ can also be viewed as a mathematical convenience. It allows some of the versatility of extended-body effects to be explored while avoiding many of the complications which would arise without it. Although it would be interesting to also explore spin and torque effects, these are left for later work.

\subsection{Torque-free bodies}
\label{Sect:Nospin}

The motion of an extended body simplifies considerably when no torque acts upon it.  Restricting to the quadrupole approximation, it follows from \eqref{quadN} that this occurs if and only if the quadrupole moment is such that
\begin{equation}
	J_1 = J_3 = 0.
	\label{torqueFree}
\end{equation}
Equivalently, the torque vanishes if and only if the conformal Killing-Yano tensor $\CKY^{ab}$ is an eigenbivector of the quadrupole moment:
\begin{equation}
	J^{ab}{}_{cd} \CKY^{cd} = - 4 J_2 \CKY^{ab}.
	\label{torqueFreeJ}
\end{equation}
These results hold in any vacuum type D spacetime. Specializing to the Schwarzschild case while recalling \eqref{quadF}, the force on a torque-free body is simply
\begin{equation}
	F_a = - \jS \nabla_a \PSi,
	\label{torqueFreeForce}
\end{equation}
where it is convenient to define
\begin{equation}
	\jS \equiv 8 \Re J_2.
	\label{jSDef}
\end{equation}
The motion is therefore affected by only a single real quadrupole component: $\jS$. Like all components of the quadrupole moment, this is unconstrained by stress-energy conservation. It may  be viewed as a kind of control parameter for suitably-engineered spacecraft.

In terms of $J^{abcd}$, it follows from \eqref{CKYKerr} and \eqref{Ji} that the control parameter can be written as
 \begin{equation}
 	\jS = -\tfrac{1}{2} \Re( J^{abcd} \CKY_{ab} \CKY_{cd}/r^2 ).
 \end{equation}
Similarly, the torque-free condition \eqref{torqueFree} is equivalent to the two complex equations
\begin{equation}
	J^{abcd} (X_{ab} \pm Y_{ab} ) \CKY_{cd} = 0.
\end{equation}
By substituting \eqref{Jexpand} into these equations, they can be rewritten in terms of the mass and momentum quadrupoles $Q^{ab}$ and $\Pi^{ab}$. In general, the resulting expressions are complicated. However, in the Newtonian limit where $\Pi^{ab} \to 0$, $r/M \to \infty$, and $\tau^a \to \partial_t$, they reduce to 
\begin{equation}
	\jS = \tfrac{3}{2} Q_{rr}, \qquad Q_{r\theta} = Q_{r \phi} = 0.
\end{equation}
This is equivalent to stating that in the Newtonian limit, $\partial_r$ must be an eigenvector of $Q^{a}{}_{b}$, and that the corresponding eigenvalue is $\frac{2}{3} \jS$. These results agree with the purely-Newtonian analysis in \cite{HarteNewtonian}, where the control parameter denoted there by $q$ is equivalent to our $\frac{2}{3} \jS$.

\subsubsection{Spin-free bodies}

We would like to consider bodies which are not only torque-free, but also spin-free\footnote{The spin-free and torque-free assumptions are logically independent. Clearly, a torque-free body can spin. More interestingly, some torqued bodies need not spin. Adopting the center of mass condition \eqref{SSC}, it follows from \eqref{Dixon} and \eqref{mvRelation} that torques satisfying $N_{ab} = 2 u_{[a} N_{b]c} u^c$ cannot spin up an initially non-spinning body. Such a torque would instead influence the dynamics via the hidden momentum $p^a - m \dot{z}_s^a = N^{a}{}_{b} u^b$. \label{Foot:torque}}. It is therefore necessary to ensure that if the torque vanishes, an angular momentum which is initially zero will remain zero. This is not automatically the case, as it follows from \eqref{Dixon} that for a torque-free body, $\dot{S}^{ab} = 2 p^{[a} \dot{z}_s^{b]}$. The right-hand side of this equation is present even in Newtonian physics \cite{HarteReview}, where it can be  eliminated by placing the origin at the center of mass. A similar strategy is effective also in the relativistic context: Imposing the center of mass condition \eqref{SSC} here and in the remainder of this section, the momentum-velocity relation \eqref{mvRelation} implies that in the torque-free case, there is some $\alpha_c$ such that $\dot{S}^{ab} = p^{[a} S^{b]c} \alpha_c$. One solution is $S^{ab} =0$. It is therefore consistent to consider non-spinning bodies when $N^{ab} = 0$ and $S^{ab} p_b = 0$. 

The spin-free, torque-free condition [and the center of mass condition \eqref{SSC}] are assumed throughout the remainder of this section. It then follows from \eqref{Dixon} and \eqref{torqueFreeForce} that 
\begin{equation}
	\dot{p}_a = - \jS \nabla_a \PSi.
	\label{pDotSimp}
\end{equation}
Also, the parameter normalization \eqref{sDef} and the momentum-velocity relation \eqref{mvRelation} imply that $p^a = \calM u^a = \calM \dot{z}^a_s$. The hidden momentum therefore vanishes and $s$ is a proper time. Furthermore, the energy \eqref{energy} and the angular momentum \eqref{Jcons} reduce to
\begin{equation}
	E = - p_a t^a, \qquad \vec{L} = \vec{z} \times \vec{p}.
\end{equation}
These quantities are constant, regardless of $\jS$.

\subsubsection{Effective potentials and effective masses}

The motion of torque-free, spin-free bodies can be described using an effective potential. To derive this, first note that the unchanging direction of $\vec{L}$ implies that the motion must be planar. Without loss of generality, we therefore restrict to the $\theta = \pi/2$ equatorial plane. Applying the conservation of $L \equiv |\vec{L}|$ then determines the azimuthal motion in terms of the radial motion:
\begin{equation}
	\frac{ d \phi }{ d s } = \frac{L}{\calM r^2}.
	\label{phiDot}
\end{equation}
Furthermore, the radial motion follows by combining this with $g_{ab} \dot{z}_s^a \dot{z}^b_s = -1$ and with the conservation of energy. The result is most conveniently expressed using a time parameter $s'$ which is related to $s$ via 
\begin{equation}
	\frac{ds'}{ ds } = M/\calM.
	\label{tauDef}
\end{equation}
In terms of this, 
\begin{equation}
	\left( \frac{dr}{ds'} \right)^2 + \Phi_\mathrm{eff} (r, \calM) = (E/M)^2,
	\label{rDot}
\end{equation}
where
\begin{align}
	\Phi_{\mathrm{eff}} (r,\calM ) \equiv (\calM/M)^2 (1-2M/r) \left[ 1 + (L/\calM r)^2 \right] 
	\label{Veff}
\end{align}
is the effective potential. Also, $d \phi / d s' = L/M r^2$.  Except for some unconventional rescalings, these equations are superficially identical to textbook results \cite{Wald} on geodesics in the Schwarzschild spacetime. They are more general, however. This is because, for geodesics, $\calM$ is fixed. For extended bodies, $m$ can vary.

In fact, if precession effects are excluded, \textit{all nontrivial aspects of the spin-free, torque-free dynamics are determined by variations in the mass}.  How $m$ varies follows from \eqref{pDotSimp}, which implies that 
\begin{equation}
	\frac{ d }{ ds } (\calM - \jS \PSi) = -\PSi \frac{d}{ds} \jS.
	\label{muDot}
\end{equation}
This suggests that it is useful to define the  ``effective mass''
\begin{equation}
	\calM_\mathrm{eff} \equiv \calM - \jS \PSi = \calM +  \jS M / r^3 .
	\label{muEff}
\end{equation}
In general, both $\calM$ and $\calM_\mathrm{eff}$ depend on time. Although either mass determines the other, it can be convenient to keep both in mind:  $m_\mathrm{eff}$ tends to simplify calculations while $m$ has a simpler interpretation. More specifically, $m$ is an ``osculating mass,'' meaning that the osculating geodesic has specific energy $E/m$ and specific angular momentum $L/m$. If $\jS$ is initially nonzero but is then rapidly reduced to zero, the subsequent motion will be a geodesic with those parameters. The physical distinction between $m$ and $m_\mathrm{eff}$ may be clarified by noting that the Newtonian potential energy due to a body's quadrupole moment is $-\! \jS M/r^3$ \cite{HarteNewtonian, Dixon70a}. The difference $\calM - \calM_\mathrm{eff}$ may therefore be interpreted as the gravitational potential energy due to a body's quadrupole moment. 

\begin{figure}
	\includegraphics[width=.97\linewidth]{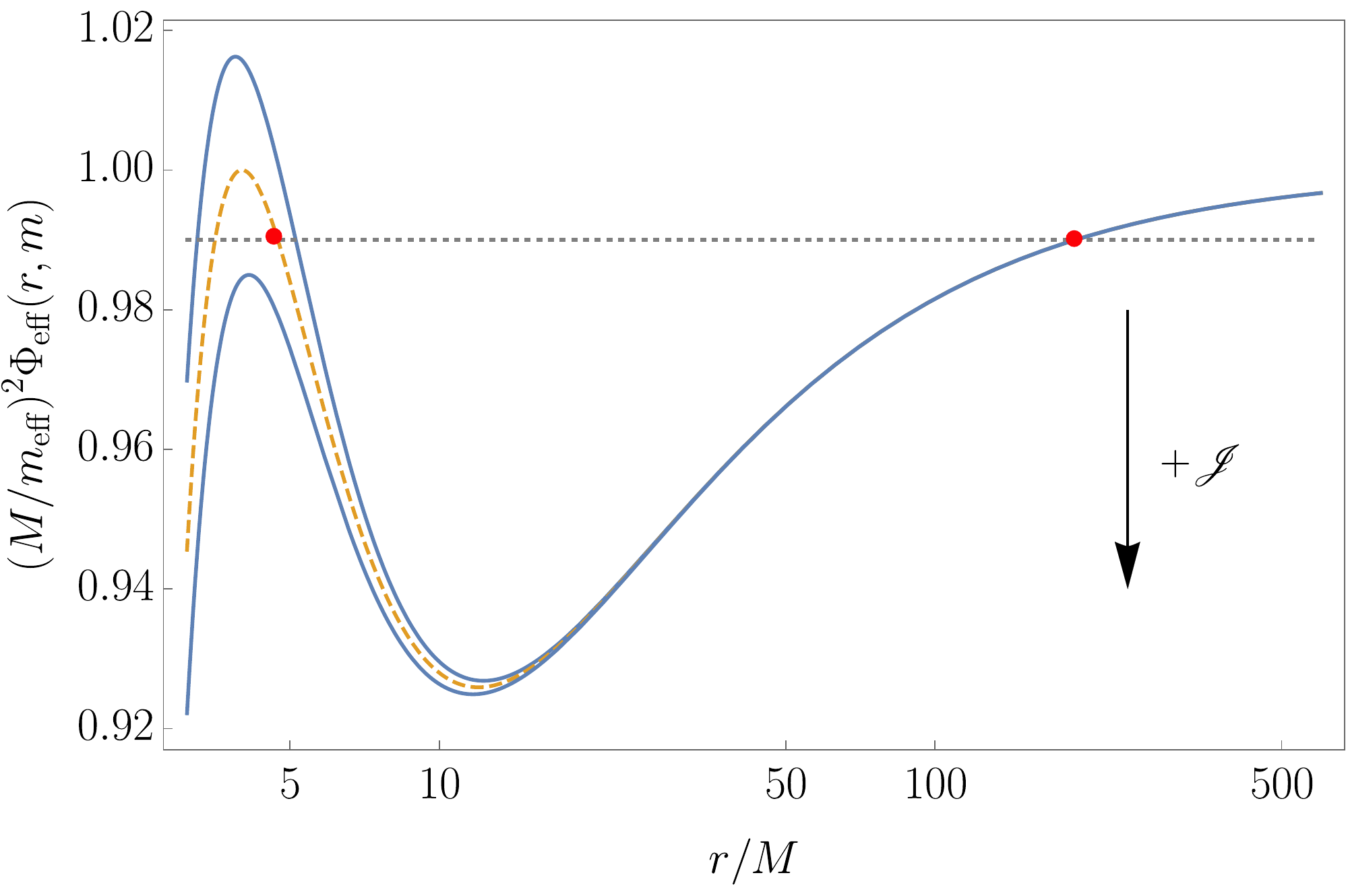}
	\caption{Effective potential energies with constant $\jS$ and $L = 4 \calM_\mathrm{eff} M$. The mass $\calM$ is given here by \eqref{muEff} with $\calM_\mathrm{eff} = \mathrm{constant}$. The dashed curve represents the $\jS = 0$ point-particle case. For the two solid curves, $\jS = \pm 2 \calM_\mathrm{eff} M^2$, where the curve with positive $\jS$ is lower. The horizontal line corresponds to $(E/\calM_\mathrm{eff})^2 = 0.99$, and the red markers are the turning points for a point-particle orbit with that energy. An extended body with this energy and with $\jS = -2 \calM_\mathrm{eff} M^2$ would have a larger  radius of pericenter. A body with this energy and with $\jS = +2 \calM_\mathrm{eff} M^2$ would have no pericenter at all; it would plunge into the central mass. Note however that these comparisons are not to be interpreted as applying to a single object which switches between different values of $\jS$. As described by \eqref{dmu}, changing configurations in this way would change $E/\calM_\mathrm{eff}$.}
	\label{Fig:PhieffConstj}
\end{figure}

Although our primary interest is in cases where $\jS$ varies, it is instructive to first suppose that it is fixed. In that case, \eqref{muDot} implies that $\calM_\mathrm{eff}$ is fixed as well. This is a special case of the more general result that $\calM - \tfrac{1}{6} J^{abcd} R_{abcd}$ is constant when $(D^\mathrm{M} J^{abcd} /ds) R_{abcd} = 0$, where the operator $D^\mathrm{M}/ds$ is a generalized Fermi derivative along $\calZ$ \cite{Dixon70a}. Regardless, the qualitative features of orbits with constant $\jS$ can be read off just by plotting $\Phi_\mathrm{eff}(r,m) = \Phi_\mathrm{eff}(r,\calM_\mathrm{eff} + \jS \PSi)$. Three such curves are displayed in Fig. \ref{Fig:PhieffConstj}.

Generalizing slightly, effective potential plots can be used in standard ways whenever $m$ depends only on $r$. This includes cases where $\jS$ is constant, although there are other possibilities as well. Regardless, if $m$ depends only on $r$, orbits are effectively fixed. Except in the context of stabilization---cf. Sect. \ref{Sect:Stab} below---such cases are not particularly interesting. Much more dramatic extended-body effects arise when $m$ changes secularly over time, which can be arranged by appropriately cycling $\jS$ over many orbits. This is discussed in Sect. \ref{Sect:boundOrb} below.

\subsubsection{Piecewise-constant $\jS$}
\label{Sect:Piecewise}

In most of the applications considered below, $\jS$ is piecewise constant. This models an idealized spacecraft which has been engineered to rapidly switch between two or more torque-free configurations. Although objects cannot switch states instantaneously without violating energy conditions, this aspect of the idealization does not appear to be essential: Finite switching times result in more complicated calculations but similar conclusions. 

To understand what happens when $\jS$ is piecewise constant, it suffices to understand how this parameter affects $m$ or $\calM_\mathrm{eff}$. Between state changes, $\calM_\mathrm{eff}$ is fixed while $\calM$ is not. During a state change, \eqref{muDot} implies that if $\jS$ rapidly switches to $\jS + \delta \! \jS$,  
\begin{equation}
	\delta \calM = 0 , \qquad \delta \calM_\mathrm{eff} = - \PSi ~ \delta \! \jS.
	\label{dmu}
\end{equation}
This result is used below to construct control strategies where changes in $\jS$ have  prescribed consequences.

Similar results arise also in Newtonian gravity. However, the interpretation is different. In a relativistic context, $E$ is constant while $m$ and $m_\mathrm{eff}$ are not. In a Newtonian context, the mass is constant while the energy is not. Despite appearances, these statements are consistent. It is only definitions which differ. In fact, there are two definitions of energy used in the Newtonian analysis in \cite{HarteNewtonian}. One of these behaves like $E - m$ and the other  like $E - m_\mathrm{eff}$.

\subsection{Radial infall}

The simplest type of motion which might be considered is radial infall. In that case, there is one interesting question: Can a mass use extended-body effects to slow or accelerate its fall? This question has been addressed before \cite{Gueron2006, Gueron2007, PoissonSwim, Vesely2019, VeselyThesis}, using constrained Lagrangians which were purported to describe cyclically-deforming spacecraft. Here, we show---without introducing interior models---that non-spinning, torque-free spacecraft can indeed control their falls. However, in contrast with some claims in the literature, no control remains if a body's configuration is cycled at high frequencies.

Suppose that an object begins at rest with initial mass $m_i$ and initial radius $r_i$. Then \eqref{tauDef}--\eqref{Veff} imply that after falling to a smaller radius $r_f < r_i$, the radial speed is given by
\begin{equation}
	\left( \frac{d r_f }{ds}\right)^2 =  
	\left( \frac{m_i}{ m_f} \right)^2 (1-2M/r_i ) - (1-2M/r_f ) .
	\label{radVel}
\end{equation}
The speed of a fall can therefore be controlled by controlling $m_f/m_i$. If a body increases its mass, it would fall more slowly than a geodesic with the same initial and final radii. If it decreases its mass, it would fall more quickly.

Whether the mass increases or decreases, and by how much, is determined by $\jS$. From \eqref{muDot}, 
\begin{equation}
	m_f / m_i = 1 - \frac{ 3 M }{ m_i } \int_{r_f}^{r_i} \! \left( \frac{ \jS }{ R^4 } \right)  dR.
	\label{massInt}
\end{equation}
This ratio is as small as possible---which maximizes the speed of a fall---when $\jS$ is as large as possible for as long as possible. Similarly, the speed of a fall can be minimized by making $\jS$ as small as possible for as long as possible. If $\jS$  can vary only in a fixed interval $[\jS_-, \jS_+]$, the largest effects are therefore obtained by holding this parameter constant at its maximum or minimum value. In those cases,
\begin{equation}
	\calM_f/\calM_i = 1 - (\jS_\pm M/m_i) \left( 1/r_f^3 - 1/r_i^3 \right) .
\end{equation}
Falls are accelerated when $\jS_+ > 0$ and slowed when $\jS_- < 0$. However,  extremely large quadrupole moments would be required to have a significant effect.

\subsection{Orbital stabilization}
\label{Sect:Stab}

We now consider circular motion, showing that extended-body effects can be used to stabilize the unstable circular geodesics with radii between $3M$ and $6M$. Again suppose that $\jS$ can vary throughout the fixed interval $[\jS_-, \jS_+]$. A nearly-circular orbit can then be stabilized by switching $\jS$ between $\jS_-$ and $\jS_+$ when $r$ crosses a fixed setpoint $\rc$. More precisely, suppose that\footnote{The discussion here is chosen to be as simple as possible, using a ``bang-bang'' control strategy without hysteresis. Many refinements are possible.}
\begin{equation}
	\jS = \begin{cases}
		\jS_+ , &	r > \rc,\\
		\jS_-, & r \leq \rc.
	\end{cases}
	\label{jStab}
\end{equation}
We assume that the orbit is not precisely circular with radius $r_c$. Then, if $r$ increases above $\rc$, it follows from \eqref{dmu} that $\calM_\mathrm{eff}$ increases by $(\jS_+ - \jS_-) M /\rc^3$. If $r$ decreases below $\rc$,  the effective mass  decreases by this same amount. As $\calM_\mathrm{eff}$ is related to $m$ via \eqref{muEff}, there must be a constant $\calM_c$ such that for \textit{all} $r$,
\begin{align}
	\calM = \calM_c - \jS  M\left(  1 / r^3 -  1 / \rc^3 \right) .
	\label{muStab}
\end{align}
This depends only on the current value of $r$, and not on the body's history. It may be substituted into the effective potential \eqref{Veff} in order to determine the behavior of orbits under the control law \eqref{jStab}.

\begin{figure}
	\includegraphics[width=\linewidth]{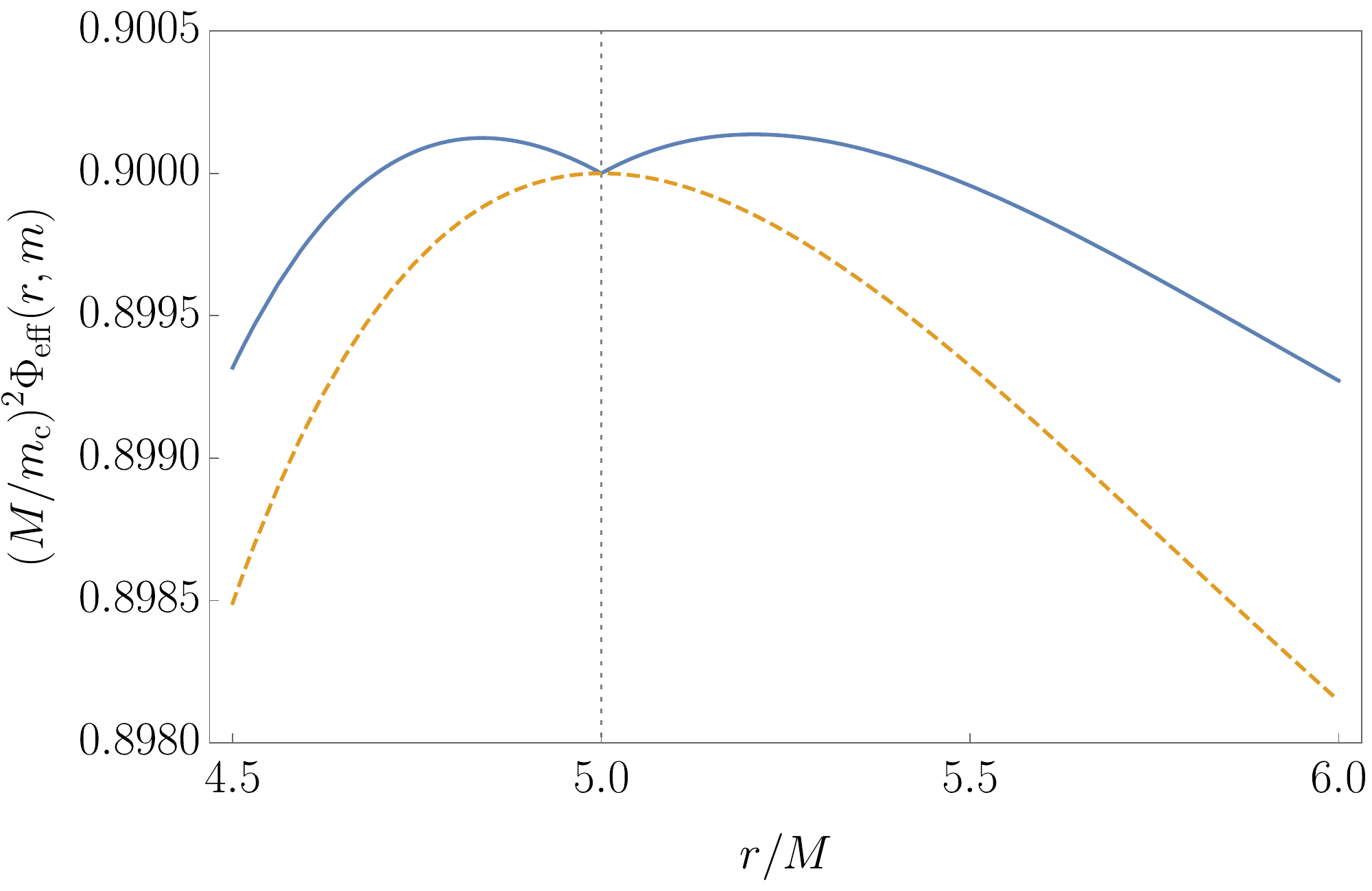}
	
	\caption{Stabilization of the unstable geodesic at $\rc = 5 M$. The mass $m$ is given here  by \eqref{muStab}. The dashed orange curve denotes the point-particle effective potential for which $\jS_\pm = 0$. The effective potential associated with the solid blue curve applies for an extended body with $\jS_\pm = \pm \calM_c M^2/4$, where the body switches between these values when crossing $r = \rc$. Extended-body effects are seen to stabilize the ordinarily-unstable circular orbit at $r = \rc$.} 
	
	\label{Fig:Stab}
\end{figure}

The result is not interesting unless $\rc$ is chosen appropriately. If the body has angular momentum $L$, we now set this radius to be equal to the radius of an unstable circular geodesic with specific angular momentum $L/\calM_c$. The setpoint $r_c$ is therefore chosen to satisfy
\begin{equation}
	L/\calM_c = \frac{ \rc }{ \sqrt{\rc/M - 3} }.
\end{equation}
Use of this and \eqref{muStab} results in the effective potential plotted in Fig. \ref{Fig:Stab}. In the point-particle case where $\jS_\pm = 0$, the effective potential has a local maximum at $r = \rc$; orbits near that radius are unstable. For an extended body in which $\jS_+$ and $\jS_-$ have opposite signs, there is instead a local minimum at $r = \rc$; orbits near that radius are stable. 

\subsection{Orbital maneuvering}
\label{Sect:boundOrb}

Extended-body effects may be used not only to stabilize a given geodesic, but also to move from one geodesic to another. This is because, although each individual orbit is close to a geodesic when $\jS$ is small, the approximating geodesic can change considerably over many orbits. There are restrictions, however. We now determine which orbital maneuvers are available to spin-free, torque-free bodies. In this context, orbital changes are determined entirely by changes in $m$. Our discussion begins without specifying precisely how mass changes occur. Later, we examine a specific strategy for obtaining large mass changes from small, cyclic variations in $\jS$. 

\subsubsection{Motion through the space of geodesics}

If a body changes its mass, its orbit may be characterized by the geodesic which instantaneously approximates it. Over time, however, the approximating (or osculating) geodesic may be viewed as drifting through the space of all possible geodesics. In order to describe this, it is convenient to introduce coordinates on the space of geodesics. Two such coordinate systems are considered here. 

Our first set of coordinates on the space of geodesics---really geodesics modulo rotations and time translations---is comprised of the specific energy $E/m$ and the specific angular momentum $L/m$. If an extended body changes its mass, the conservation of $E$ and $L$ implies that the geodesics which instantaneously approximate its orbit must be confined to a straight line in the coordinates $(E/m,L/m)$. Different mass-changing bodies can move on different lines, each of which is characterized by the slope $L/E$. Mass increases move the approximating geodesics towards the $E/m = L/m = 0$ origin. Mass decreases move them away. Although these statements are simple to derive, their physical implications are not immediately apparent.

It is more intuitive to describe a body's motion using a different set of coordinates on the space of geodesics: the eccentricity $e$ and semi-latus rectum $p$. These quantities generalize the eccentricity and semi-latus rectum which appear in the Newtonian 2-body problem\footnote{For a Keplerian ellipse with semi-major axis $a$ and eccentricity $e$, the semi-latus rectum is $a(1-e^2)$. Geometrically, this is the half-length of the chord which passes through a focus and is orthogonal to the major axis of the ellipse. The Newtonian semi-latus rectum is also proportional to the square of the angular momentum of the orbiting body.}. Their precise definitions are given in Appendix \ref{App:epDef} and elsewhere \cite{CutlerGeodesics, BarackSago, Chandra}, although the idea is that $e$ and $p$ are defined to preserve, up a non-dimensionalization of $p$, the Keplerian relations between themselves, the radius of apocenter $r_+$, and the radius of pericenter $r_-$. This means that
\begin{equation}
		r_\pm = \frac{ p M }{ 1 \mp e }.
		\label{rPM}
\end{equation}
Equivalently,
\begin{equation}
	p M = \frac{ 2r_+ r_-}{ r_+ + r_- }, \qquad e = \frac{ r_+ - r_- }{ r_+ + r_- } .
	\label{peDef}
\end{equation}
As explained in Appendix \ref{App:epDef}, these relations hold throughout the region of parameter space described by
\begin{equation}
	p \geq 6 + 2e , \qquad 0 \leq e < 1.
	\label{separatrix}
\end{equation}
If $e = 0$, a geodesic is circular and stable, with radius $pM$. If $e \to 1$, a geodesic extends to arbitrarily large radii; it is unbound. If $p = 6+2e$, the geodesic asymptotically approaches an unstable circular geodesic with radius $2pM/(p-4)$. For the remainder of this section, we use $(e,p)$ as coordinates on the space of geodesics. The coordinates $(E/m,L/m)$ are related via \eqref{ELpe}.

\begin{figure}
	\includegraphics[width=\linewidth]{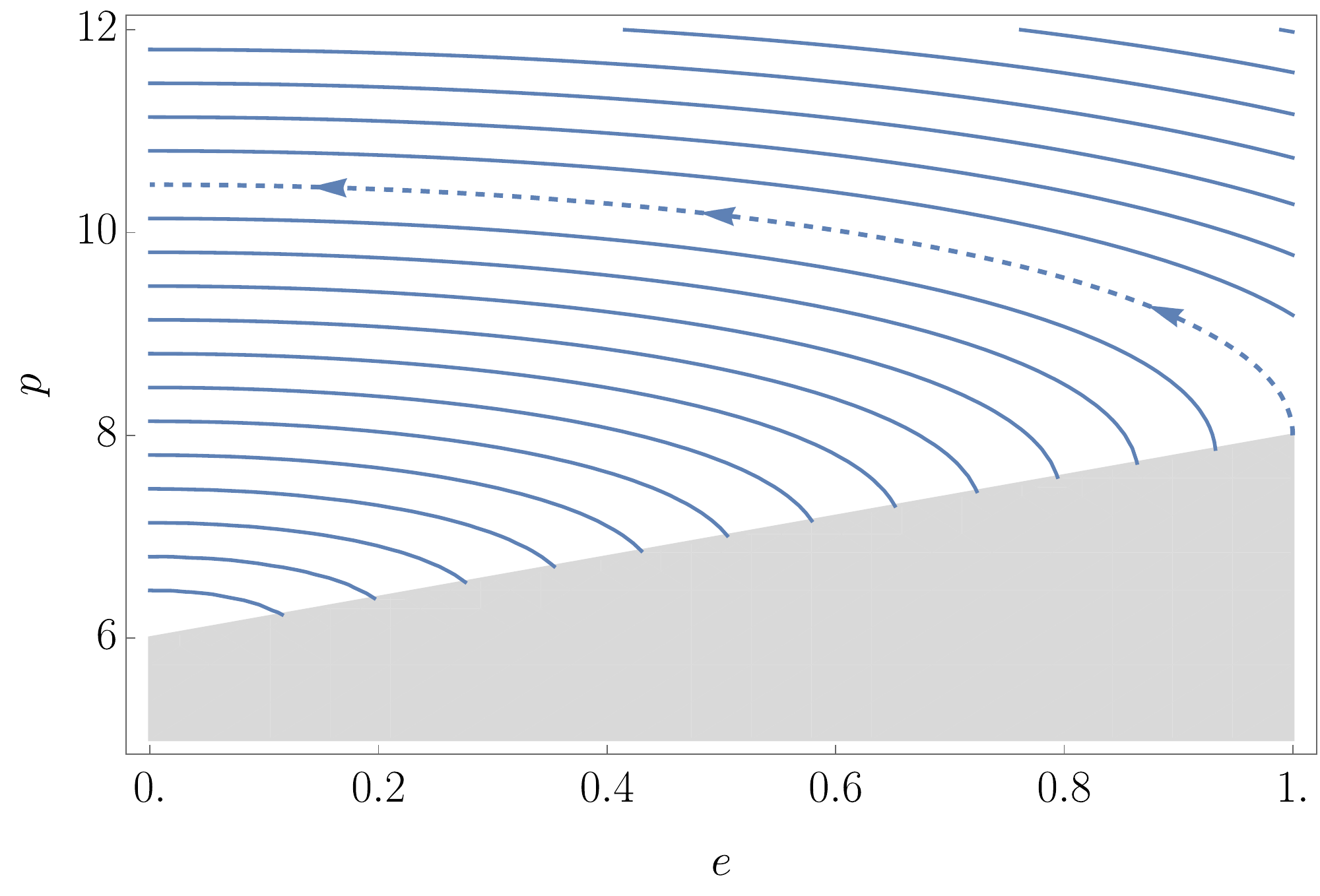}
	\caption{Trajectories through the space of geodesics when $m$ is slowly varied. Geodesics are labelled by their semi-latus rectum $p$ and eccentricity $e$. The shaded portion of the plot is the region excluded by \eqref{separatrix}. As indicated by the arrows, moving along any curve from right to left corresponds to a body increasing its mass. Points on the vertical axis correspond to stable circular orbits with radii $pM$. For curves above the dashed line, decreasing mass eventually results in an unbound orbit. For curves below the dashed line, decreasing mass eventually results in an intersection with the separatrix. Geodesics on the separatrix asymptotically approach unstable circular orbits, even when $e \neq 0$.} 
	\label{Fig:pePlot}
\end{figure}

For a torque-free extended body in Newtonian physics, extended-body effects can be used to change $e$ but not $p$ \cite{HarteNewtonian}. We now show that in a relativistic context, changes in $m$ change both $e$ and $p$. To see this, note that the conservation of $E$ and $L$ implies that the geodesics which approximate the motion of a mass-changing body must lie on a level curve of
\begin{equation}
		(E/L)^2 = \frac{ (p-2)^2 -4e^2 }{ M^2 p^3 },
		\label{ELratio}
\end{equation} 
viewed as a function of $e$ and $p$. A number of these curves are plotted in Fig. \ref{Fig:pePlot}. Crucially, they can be separated into two categories, depending on whether or not they extend to the $p=6+2e$ separatrix. If a level curve does not extend to the separatrix, an initially-bound geodesic can eventually be made unbound. If a level curve does extend to the separatrix, escape is impossible.

If escape \textit{is} possible, \eqref{separatrix} implies that $p > 8$ as $e \to 1$. Using \eqref{ELratio}, this condition can be satisfied only when
\begin{equation}
	EM/L < 1/4.		
	\label{EBound1}
\end{equation}
A more intuitive characterization may be obtained by supposing that an orbit is initially circular, with radius $r_i$. Escape is then possible for all
\begin{equation}
	r_i > 2(3+ \sqrt{5})M \approx 10.5M.
	\label{rcMin}
\end{equation}
Transferring from a stable circular orbit to an unbound orbit is accomplished by decreasing $m$. From \eqref{ELpe}, the initial and final masses must be related by
\begin{equation}
	m_f/m_i = \frac{ 1-2M/r_i }{ \sqrt{ 1-3M/r_i } } .
\end{equation}
This ratio is smallest when $r_i$ saturates the bound \eqref{rcMin}, in which case $m_f/m_i \approx 0.958$. It may also be shown that in a post-Newtonian limit where $p \to \infty$, escaping from an initially-circular orbit with $p_i = r_i/M$ results in the semi-latus rectum $p_f =p_i -4/p_i + \ldots$ The first deviation from the constant-$p$ Newtonian expectation therefore occurs at second post-Newtonian order.

If $E M/L \geq  1/4$, a body cannot use mass changes to escape from a bound orbit. Instead, a decreasing mass would eventually result in the approximating geodesic approaching the separatrix. If this approach is not made with care, instability would set in and the body would plunge. However, extended-body effects can be used to stabilize the approach. In that case, the orbit would eventually tend towards an unstable circular geodesic. In other words, it is possible---in the strong-field regime---to use mass changes to transfer between pairs of circular orbits. The initial and final radii of these orbits cannot be chosen at will, but must be related by
\begin{equation}
	\left( \frac{ 1-2M/r_i }{  1 -2M/r_f } \right)^2  = \frac{ r_i }{ r_f }.
\end{equation}
Transitions are therefore possible between circular geodesics with radii between $4M$ and $2(3+ \sqrt{5})M$. Except in the degenerate case where $r_i = r_f = 6M$, this transition always occurs from a stable geodesic to an unstable one, or vice versa. Furthermore, the mass change required to perform such a maneuver is described by
\begin{equation}
	(m_f/m_i)^2 = \frac{ r_i }{ r_f } \left( \frac{ 1 - 3M/r_f }{ 1-3M/r_i } \right).
\end{equation}

\begin{figure}
	\includegraphics[width=.98\linewidth]{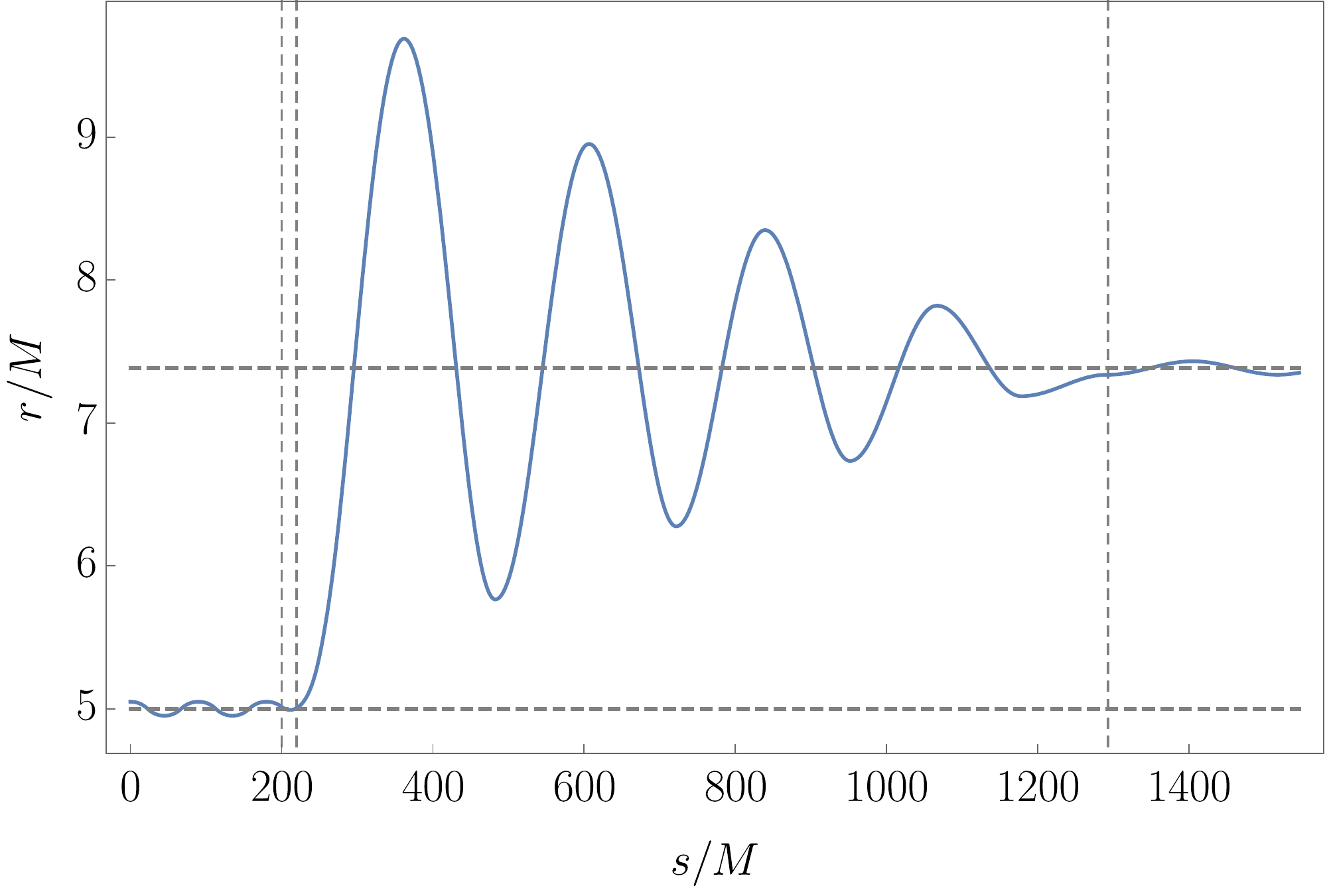}
	\caption{Transition from one circular orbit to another. The motion begins in a nearly-circular orbit with radius $5.0M$. It ends in a nearly-circular orbit with radius $7.4M$. There are four control regimes, with boundaries marked by vertical lines. Initially, the orbit is stabilized by switching $\jS$ according to \eqref{jStab}. Next, the body is destabilized towards larger radii by briefly setting $\jS=\jS_-$, which places it in a nearly-geodesic orbit with $p= 6.7$ and $e=0.33$. Between $s = 220M$ and $s = 1290M$, $\jS$ is switched according to \eqref{jMass}. This increases $\calM$ and decreases $e$. Finally, all extended-body effects are switched off when $s > 1290M$. In all phases except the last, $\jS_\pm = \pm m_i M^2/8$, where $m_i$ denotes the initial value for $m$. This plot was obtained by numerically integrating \eqref{rDot} and \eqref{muDot}. }
	\label{Fig:circTransition}
\end{figure}

An example of an extended body maneuvering between two circular geodesics is provided in Fig. \ref{Fig:circTransition}. The body considered there begins near an unstable geodesic with radius $5.0M$, which is initially stabilized by the method discussed in Sect. \ref{Sect:Stab}. Using the mass-increasing technique discussed below, this is then converted into a stable circular geodesic with radius $7.4M$. The process requires a total mass increase of approximately $0.25\%$. Note that although the initial and final geodesics are nearly circular, the intermediate geodesics are not.

\subsubsection{Changing mass}

We now explain how it is that large changes in mass can be produced without correspondingly large quadrupole moments. First, it is clear from \eqref{muDot} and \eqref{muEff} that changes in $m_\mathrm{eff}$ are produced by changes in $\jS$. However, if $\jS$ changes and then returns to its original value, it is not necessarily true that $m_\mathrm{eff}$ also returns to its original value. In fact, appropriately-designed cycles can result in net increases or decreases in $m_\mathrm{eff}$. Although each such change may be small, a body in a bound orbit can complete many cycles, eventually resulting in a significant mass change. 

\begin{figure}
	\includegraphics[width=1\linewidth]{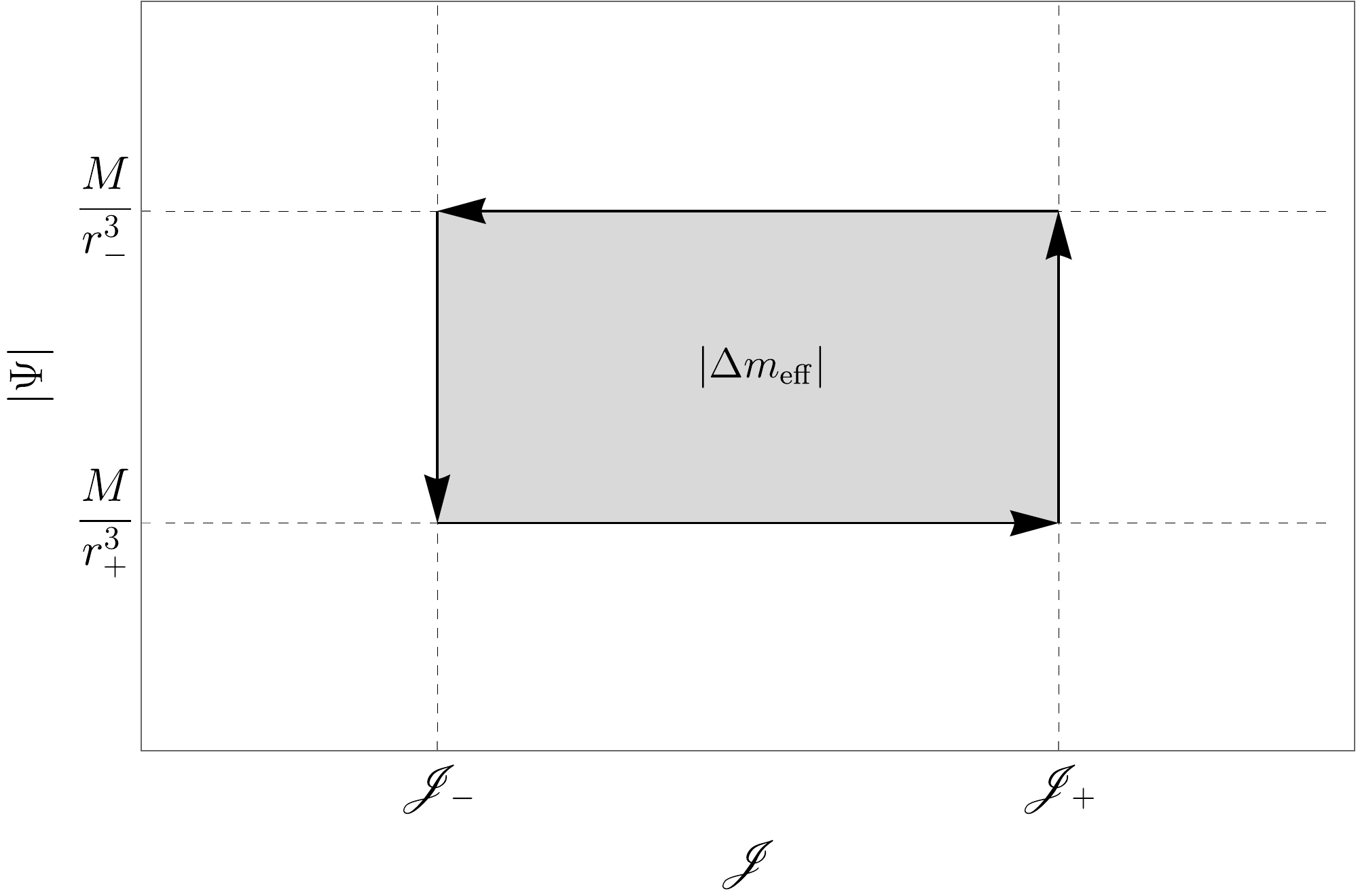}
	\caption{The mass-decreasing cycle described by \eqref{jMass}. The area enclosed by the cycle is $|\Delta m_\mathrm{eff}|$. The arrows indicate the direction of increasing time. If such a cycle were repeated many times, the area here would slowly change as  $r_+$ and $r_-$ evolve in response to the changing mass.}
	\label{Fig:cycle}
\end{figure}

A single cycle may be viewed as a closed curve in the space parametrized by $\PSi$ and $\jS$. It follows from \eqref{muDot} and \eqref{muEff} that the magnitude of the change $\Delta m_\mathrm{eff}$ in the effective mass is equal to the area enclosed by that curve. If $\jS$ is again allowed to vary throughout the interval $[\jS_-, \jS_+]$, a body in a non-circular orbit can optimally change its mass by choosing the rectangular cycle illustrated in Fig. \ref{Fig:cycle}. This corresponds to setting
\begin{equation}
	\jS = \begin{cases}
		\jS_- , &	\dot{r} > 0, \\
		\jS_+ , & \dot{r} \leq 0,
	\end{cases}
	\label{jMass}
\end{equation}
which decreases the mass. A body can instead increase its mass by swapping the roles of $\jS_-$ and $ \jS_+$ in this expression, which corresponds to reversing the directions of the arrows in Fig. \ref{Fig:cycle}. With either of these strategies, $\jS$ is piecewise-constant and the effective mass changes only when passing through pericenter or apocenter, where it jumps according to \eqref{dmu}. Applying the mass-decreasing strategy over one orbit, $\jS$ returns to its original value while
\begin{align}
	\Delta m_\mathrm{eff} &= -(\jS_+ - \jS_-)M \left( 1/r_-^3 - 1/r_+^3 \right)
	\nonumber
	\\
	&= -2 e (\jS_+ - \jS_-) \left( \frac{ 3+e^2 }{ p^3 M^2 } \right).
	\label{dmeff}
\end{align}
Repeating such a cycle over multiple orbits would result in a continually decreasing mass, at least until $e\to 1$ or $p \to 6 + 2e$. If a mass-increasing cycle were used instead, the mass would continue to increase until the orbit circularizes.

Changes in $m_\mathrm{eff}$ induce changes in the eccentricity $e$ and the semi-latus rectum $p$. In the mass-decreasing case, \eqref{dmeff} and \eqref{ELpe} imply that over each orbit,
\begin{align}
	\Delta e = &~ 2 \left( \frac{ \jS_+ - \jS_- }{ L^2/m  } \right) 
	\nonumber
	\\
	& ~ \times \frac{ (3+e^2) [p-8+(12/p)(1-e^2)] }{ (p-6)^2-4e^2 }  ,
\end{align}
and
\begin{align}
	\Delta p = -16 \left( \frac{ \jS_+ - \jS_- }{ L^2/m  } \right) \frac{ e (3+e^2) }{ (p-6)^2 - 4 e^2 } .
\end{align}
The denominators in both of these expressions vanish on the separatrix, which is a consequence of the instability which sets in as $p \to 6+2e$. Another implication of these expressions is that even though $\Delta m_\mathrm{eff}$ and $\Delta p$ both vanish as $e \to 0$, the same is not true for $\Delta e$. It is therefore possible for a body to increase its eccentricity even if its orbit is initially circular. This is because a circular orbit can be made slightly eccentric by changing $\jS$, and this small eccentricity may be used to build up a larger eccentricity---and eventually an appreciable mass change.

Lastly, we note that changes in $m_\mathrm{eff}$, $p$, and $e$ can be seen to depend on $\jS_\pm$ only via the difference $\jS_+ - \jS_-$. However, this is not the only measure of the quadrupole moment which affects the dynamics. Without entering into details, the sum $\jS_+ + \jS_-$ controls the rate of orbital precession. It has a similar effect also in Newtonian gravity \cite{HarteNewtonian}.

\section{Discussion}
\label{Sect:Discuss}

Our results may be grouped into two categories. First, in Sect. \ref{Sect:TypeD}, we derived general characteristics associated with the laws of motion in vacuum type D spacetimes. Second, in Sect. \ref{Sect:Schw}, we applied those characteristics to understand the behavior of actively-controlled spacecraft in the Schwarzschild spacetime. We now discuss these results separately.

\subsection{Constraints on the laws of motion}

Working in vacuum type D spacetimes and in the quadrupole approximation, we have shown that all gravitational forces and torques can be summarized by \eqref{genForce}, or equivalently by \eqref{quadF} and \eqref{quadN}. These results depend on the background geometry only via Lie derivatives of the Weyl scalar $\PSi$ and of the conformal Killing-Yano tensor $\CKY_{ab}$. One consequence is that at least four of a body's ten quadrupole components cannot affect its motion. Similarly, at least four force and torque combinations are impossible, regardless of a body's internal structure: Beyond the Killing constraints with the form \eqref{fnCons}, we have shown that the torque must be constrained by $N^{ab} \CKY_{ab} = 0$. Taken together, all force and torque constraints involve either Killing vectors or conformal Killing-Yano tensors. However, unlike the former of these constraints, the latter do not appear to be related to  conservation laws.

The calculations used to derive these results can easily be extended to understand how forces and torques behave in spacetimes which are not necessarily type D. For example, a straightforward calculation similar to the one which led to \eqref{genForce} shows that in the quadrupole approximation, the generalized force in any vacuum type N spacetime is
\begin{align}
	\frac{ d }{ ds } \calP_\xi = \tfrac{4}{3} \Re [ \Psi_4 ( J_0 Y^{ab} + J_1 Z^{ab}) \mathscr{L}_\xi X_{ab} 
	\nonumber
	\\
	~ - J_0 \mathscr{L}_\xi \Psi_4 ] .
	\label{typeN}
\end{align}
Here, the lone principal null direction is now tangent to $\ell^a$, so the only non-vanishing Weyl scalar is $\Psi_4 \equiv \frac{1}{4} R_{abcd} Y^{ab} Y^{cd}$. Physically, this result could be used to describe how extended bodies move in response to a gravitational wave. The absence of $J_2$, $J_3$, and $J_4$ implies that at least six real quadrupole components are irrelevant in vacuum type N spacetimes. This contrasts with the four irrelevant components found in the type D case. Similarly, there are at least six constraints on force and torque combinations in vacuum type N spacetimes. Some type N spacetimes admit only a single Killing vector \cite{GriffithsExact}, so in at least some cases, the majority of these constraints are not due to Killing fields. It may also be seen that unlike in the type D case, a spacecraft which actively controls its torque can, in a type N background, have no independent control over its force. 

\subsection{Applications to rocket-free spacecraft}

\begin{table*}
  \centering
  \setlength{\tabcolsep}{9pt}
  \caption{\label{tab:symbols} Comparison of spin-free, torque-free bodies in Schwarzschild and torque-free bodies in a spherically-symmetric Newtonian potential. Results and notation in the rightmost column are taken from \cite{HarteNewtonian}. Forces are seen to be identical if the relativistic control parameter $\jS$ is identified with the Newtonian $\tfrac{3}{2}q$. The final two rows refer to cases where the control parameter rapidly jumps between two values, as described here in Sect. \ref{Sect:Piecewise}.}
  
  \begin{tabular}{p{.35\textwidth} p{0.25\textwidth} p{10em}} \hline\hline
    Concept & Relativistic & Newtonian\\ 
    \hline
    Quadrupole moment & $J^{abcd}$	&	$Q_{ij}$
    \\
    Torque-free condition	& $J^{ab}{}_{cd} \mathcal{K}^{cd} = - 4 J_2 \mathcal{K}^{ab}$		& $Q^{i}{}_{j} r^j =q r^i$
    \\
   	Torque-free quadrupole force	& $\nabla_a (\jS M/r^{3})$ 	& $ \nabla_i ( \tfrac{3}{2} q M/r^3 )$ 
   	\\
    Torque-free control parameter	&	$\jS = 8 \Re J_2$	&	$q$
    \\
    Always constant	&	$E$, $L$	&	$m$, $L$
    \\
    Constant when control parameter is constant 	&	 $m_\mathrm{eff}$	&	 $E$ \\
    Constant when control parameter jumps	&
   		$m = m_\mathrm{eff} - \jS M/r^3$  & $E_\mathrm{pt} = E + \tfrac{3}{2} q M/r^3$ \\
   	Effect of control parameter jump &	$\delta m_\mathrm{eff} = \delta \! \jS M/r^3$ & $\delta E= - \tfrac{3}{2} \delta q M /r^3$ 
     \\ 
     \hline\hline
  \end{tabular}
  \label{TableNewt}
\end{table*}

Our second category of results involve spacecraft which actively control their quadrupole moments in order to control their motion. Perhaps surprisingly, the behavior of spin-free, torque-free bodies in Schwarzschild is almost identical to the behavior of torque-free \textit{Newtonian} bodies in spherically-symmetric gravitational fields. For example, the torque-free condition is equivalent, in both cases, to an eigenvector condition on the quadrupole moment: In the Newtonian case, it is the radial vector which must be an eigenvector of the mass quadrupole moment \cite{HarteNewtonian}. In the relativistic case, it is $\CKY^{ab}$ which must be an eigen(bi)vector of $J^{abcd}$. The eigenvalues associated with these eigenvectors  control the motion. More precisely, they control the  force, which is structurally identical in both the Newtonian and the relativistic cases. These and other characteristics of the Newtonian and relativistic problems are compared in Table \ref{TableNewt}.

There are only two significant differences between the capabilities of spin-free, torque-free spacecraft in the Newtonian and the relativistic contexts: In Schwarzschild, (i) unstable geodesics  may be stabilized, and (ii) it is possible to transfer between certain pairs of circular orbits with radii between $4M$ and $2(3+\sqrt{5})M$. In a spherically-symmetric Newtonian field, there are, by contrast, no instabilities to stabilize. It is also not possible, in the Newtonian context, to transfer from one circular orbit to another (unless the torque-free condition is lifted). These differences are not due to any intrinsically-relativistic effects in the theory of motion, but rather to the relative complexity of the Schwarzschild geometry. This is evidenced by the fact that features similar to those seen in Schwarzschild can also be found in Newtonian systems which are not spherically symmetric. For example, unstable point-particle orbits can arise around highly-oblate Newtonian masses, and these may be stabilized in a manner similar to the relativistic stabilization discussed in Sect. \ref{Sect:Stab} \cite{HarteNewtonian}.

This is not to say that extended-body effects are always similar in Newtonian and relativistic contexts. If the spin-free and torque-free assumptions made in Sect. \ref{Sect:Schw} are lifted, there can be essential differences. One such difference is that the relativistic theory allows a body to directly control its hidden mechanical momentum: Even though it follows from \eqref{Dixon} that $p^a$ cannot change on very-short timescales, \eqref{mvRelation} implies that $\dot{z}^a_s$ can vary as rapidly as the multipole moments can be varied. A body can therefore exert direct control not only over its acceleration, but also over its velocity. This is a fundamentally non-Newtonian effect. Some of its consequences have been investigated previously in cosmological backgrounds \cite{Harte2007}.

While we have not focused here on direct velocity control in black hole spacetimes---which is not possible in the spin-free, torque-free context of Sect. \ref{Sect:Schw}---some brief comments are in order. First, it follows from \eqref{mvRelation} that in an (at least instantaneously) spin-free context, the directly-controllable portion of the velocity is $\dot{z}^a_s = (\ldots)  - N^{a}{}_{b} p^b/m^2$. Furthermore, use of \eqref{NW} shows that in Schwarzschild and in the quadrupole approximation, a body whose momentum is initially aligned with $t^a$ can directly control \textit{only} its angular velocities $\dot{\theta}$ and $\dot{\phi}$. Rapidly altering $\dot{r}$ is not possible unless $L \neq 0$. A body which falls radially in Schwarzschild can have a small amount of angular momentum, and it can use this angular momentum to directly alter the speed of its fall. However, the magnitude of this effect would be quadratic in the quadrupole moment and therefore miniscule. 


\appendix

\section{Notation and conventions}
\label{App:Conv}

This paper assumes four spacetime dimensions. Sign and index conventions follow those of Wald \cite{Wald}: The metric $g_{ab}$ is assumed to have signature $+2$. Units are used in which $c=G=1$. Abstract spacetime indices are denoted by $a,b, \ldots$, spinor indices by $A,B,\ldots$, four-dimensional coordinate indices by $\mu, \nu, \ldots$, and other numerical indices by $i, j , \ldots$ The Riemann tensor $R_{abc}{}^{d}$ satisfies $2 \nabla_{[a} \nabla_{b]} \omega_c \equiv (\nabla_a \nabla_b - \nabla_b \nabla_a) \omega_c = R_{abc}{}^{d} \omega_d$ for any 1-form $\omega_a$. In some places, arrows are placed over symbols to denote triples which are analogous to Euclidean 3-vectors. Overdots are used to denote derivatives only with respect to the worldline parameter $s$, not with respect to other parameters such as $s'$. The notation $(\ldots)_\mathrm{TF}$ is used to indicate the trace-free component of the object inside the parentheses, where the trace-free operation is performed in such a way that it maintains the index symmetries and orthogonality properties of the original object. The main symbols used in the paper are summarized in Table \ref{Table}.

\begin{table*}
  \centering
  \setlength{\tabcolsep}{9pt}
  \caption{\label{tab:symbols} Table of symbols. There are  two categories: symbols associated with the spacetime geometry and symbols associated with a body moving in that geometry.}
  \begin{tabular}{p{8em} p{0.4\textwidth} p{12em}}
	\hline\hline
    Symbol & Description & Reference \\
    \hline
    $\ell^a$, $n^a$, $m^a$, $\bar{m}^a$ & Null tetrad &	\eqref{tetradProd} 
    \\
    $X_{ab}$, $Y_{ab}$, $Z_{ab}$ & Self-dual 2-form basis elements & \eqref{2formBasis}, \eqref{bivectProd} 
    \\
    $\CKY_{ab}$	& Self-dual conformal Killing-Yano tensor 	& \eqref{CKY}, \eqref{CKYKerr}, \eqref{CKYcoords}
    \\
    $\PSi$	& 	Weyl scalar more commonly denoted by $\Psi_2$		& \eqref{Rcurv}, \eqref{Psi2}
    \\
        $M$, $a$	& Kerr mass and specific angular momentum	&	\eqref{Kerr}	
    \\
    $t^a$, $\psi^a_{(i)}$	&	Killing vector fields	&	\eqref{Killing}
    \\
    $t$, $r$, $\theta$, $\phi$	&	Boyer-Lindquist coordinates	&	\eqref{Kerr} \\
    $\vec{z}$, $\vec{\theta}$, $\vec{\phi}$	&	3-vector basis in Schwarzschild	&	\eqref{thetaphiDef}, \eqref{Jcons}
    \\
    \hline
    $\calZ$, $z_s$	& Reference worldline and a point on that worldline	&	\eqref{zsDef} \\
    $\Sigma$, $\Sigma_s$ 	& Foliation $\Sigma = \{\Sigma_s|s\}$ and a leaf in that foliation	&  \eqref{zsDef}	\\ 
     $s$, $s'$	& Worldline parameters	&	\eqref{sDef}, \eqref{tauDef}\\
    $\xi^a$		&	Generalized Killing vector	&	\eqref{PDef}, \eqref{Lieg}	
    \\
    $T_{ab}$	&	Stress-energy tensor	&	\eqref{stressCons}, \eqref{PDef}
     \\ 
     $\calP_\xi$	&	Generalized momentum	&	\eqref{PDef}, \eqref{pSDef}
    \\
    $p_a$, $S^{ab}$	&	Linear and angular momenta &	\eqref{Dixon}, \eqref{pSDef}	\\
    $F_a$, $N^{ab}$	&	Force and torque		& \eqref{FNDef}, \eqref{Dixon}, \eqref{FN}	
    \\
    $\calM$, $\calM_\mathrm{eff}$		& Ordinary and effective masses	& \eqref{mDef}, \eqref{muEff}\\
    $\tilde{J}_{abcd}$	& Full quadrupole moment		&	\eqref{Jsyms}, \eqref{Jint}, \eqref{JexpandFull}
    \\
    $J_{abcd}$	&	Trace-free quadrupole moment 	&  \eqref{JTF}, \eqref{Jexpand}
    \\
    $Q^{ab}$, $\Pi^{ab}$	&	Mass and momentum quadrupole moments	&	\eqref{Qremap}, \eqref{Jexpand}
    \\
    $J_0, \ldots, J_4$		&	Analogs of Weyl scalars for $J_{abcd}$	&	\eqref{Ji}, \eqref{quad}
    \\
    $\jS$	& $8 \Re J_2$ & 	\eqref{torqueFreeForce}, \eqref{jSDef}
    \\
    $\Phi_\mathrm{eff}$	&	Effective potential	& \eqref{rDot}, \eqref{Veff}	
    \\
       $E$, $\vec{L}$	&	Energy and angular momentum	 &	\eqref{energy}, \eqref{consLaws}, \eqref{Jcons}, \eqref{ELpe} 
    \\
    $e$, $p$	&	Eccentricity and semi-latus rectum	& \eqref{rPM}, \eqref{peDef}, \eqref{ELpe} 
    \\ 
    $r_+$, $r_-$ & Radii of apocenter and pericenter	&	\eqref{rPM}	\\
     \hline\hline
  \end{tabular}
  \label{Table}
\end{table*}

\section{Kerr and Schwarzschild geometries}
\label{App:Kerr}

This appendix collects various facts and conventions related to the Kerr and Schwarzschild spacetimes. In Boyer-Lindquist coordinates $(t,r,\theta,\phi)$, the line element of a Kerr spacetime with mass $M$ and specific angular momentum $a$ is
\begin{align}
	ds^2 = - \left( 1 - \frac{ 2 M r }{ \SigmaK }\right) dt^2 - \frac{ 4 a M r }{ \SigmaK } \sin^2 \theta dt d\phi + \frac{ \SigmaK }{ \DeltaK } dr^2 
	\nonumber
	\\
	~ \SigmaK d\theta^2 + \frac{ (r^2+a^2)^2 - \DeltaK a^2 \sin^2 \theta }{ \SigmaK } \sin^2 \theta d\phi^2,
	\label{Kerr}
\end{align}
where $\DeltaK \equiv r^2 - 2 Mr +a^2$ and $\SigmaK \equiv r^2 + a^2 \cos^2 \theta$. If $a = 0$, this reduces to the Schwarzschild metric in standard coordinates.

The discussion in Sect. \ref{Sect:TypeD} makes use of a null basis $(\ell^a, n^a, m^a , \bar{m}^a)$ in which the first two elements are aligned with the spacetime's principal null directions. Much of that discussion is valid in any vacuum type D spacetime. However, in the special case of the Kerr spacetime, one example of an appropriate basis is
\begin{subequations}
\label{tetradKerr}
\begin{align}
	\sqrt{2 } \ell^a &= \frac{ 1 }{  \DeltaK } \left[ (  r^2 + a^2 ) \partial_t + \Delta_K \partial_r +  a \partial_\phi \right],
	\\
	\sqrt{2 }  n^a &= \frac{ 1 }{ \SigmaK } \left[ ( r^2 + a^2 ) \partial_t - \DeltaK \partial_r + a  \partial_\phi \right] ,
	\\
	\sqrt{2 }  m^a &=  \frac{   i a \sin \theta \partial_t  + \partial_\theta  +  i \csc \theta  \partial_\phi  }{ r + i a \cos \theta  }  .
\end{align}
\end{subequations}
With this choice, $\ell^a$ is tangent to the outgoing principal null direction. Similarly, $n^a$ is tangent to the ingoing principal null direction. Other possible tetrads can be related to this one by the rescalings \eqref{rescale} and the discrete swaps $\ell^a \leftrightarrow n^a$ and $m^a \leftrightarrow \bar{m}^a$. 

A significant role is played in the paper by the self-dual conformal Killing-Yano tensor $\CKY_{ab}$. In Kerr, this is related to the coordinates and the tetrad via \eqref{2formBasis} and \eqref{CKYKerr}. Using \eqref{tetradKerr}, its coordinate components are
\begin{align}
	\mathcal{K}_{ab} &= \tfrac{1}{2} d (t- a \phi) \wedge d \left[ i (r - i a \cos \theta)^2 \right]
	\nonumber
	\\
	& ~ +(r - i a \cos \theta)^3 d \left(  \frac{ r \cos \theta}{ r- i a \cos \theta }  \right) \wedge d\phi.
	 \label{CKYcoords}
\end{align}
Although computed using a specific tetrad, this is in fact invariant under the rescalings \eqref{rescale}. In Schwarzschild, it reduces to $\CKY_{ab} =  r [ i (dt \wedge dr) - r^2 \sin \theta (d\theta \wedge d\phi)]$.

Our final comments are concerned with Killing vectors. In the Schwarzschild spacetime, there are four independent Killing vectors, denoted here by
\begin{subequations}
\label{Killing}
\begin{gather}
	\psi^a_{(1)} \equiv - \sin \phi \partial_\theta - \cot \theta \cos \phi \partial_\phi,
	\\
	\psi^a_{(2)} \equiv \cos \phi \partial_\theta - \cot \theta \sin \phi \partial_\phi,
	\\
	\psi^a_{(3)} \equiv  \partial_\phi, \qquad t^a \equiv \partial_t. 
	\label{KillingKerr}
\end{gather}
\end{subequations}
The three $\psi^a_{(i)}$ generate rotations while $t^a$ generates time translations. In Kerr spacetimes with $a \neq 0$, only $t^a$ and $\psi^a_{(3)}$ remain Killing.

\section{Eccentricity, semi-latus rectum, and the separatrix}
\label{App:epDef}

In Sect. \ref{Sect:Schw}, two methods were used to parametrize bound timelike geodesics in the Schwarzschild spacetime. The first method parametrized an equatorial geodesic by its specific energy $E/m$ and its specific angular momentum $L/m$. The second parametrization instead describes an equatorial geodesic in terms of its eccentricity $e$ and semi-latus rectum $p$. Up to a non-dimensionalization of $p$, these parameters are defined to preserve the non-relativistic Keplerian relations between $p$, $e$, and the turning points $r_\pm$ \cite{CutlerGeodesics, BarackSago, Chandra}; cf. \eqref{rPM} and \eqref{peDef}. There are, however, subtleties. This Appendix reviews the definitions for $e$ and $p$ and shows that these parameters are constrained by the inequalities \eqref{separatrix}.

First, note that \eqref{Veff} can be used to show that for any timelike geodesic, a radial turning point $r_t$ must satisfy
\begin{align}
	0 = \rt^3 \left[ E^2 - M^2 \Phi_\mathrm{eff}( \rt, m)  \right] = (E^2 - m^2 ) \rt^3 
	\nonumber
	\\
	~ + 2 m^2 M \rt^2 - L^2 \rt + 2 L^2 M .
	\label{cubic}
\end{align}
The right-hand side here is a cubic polynomial in $r_t$, so there are three possible solutions. If two of those solutions can be written as $r_1 \equiv p M /( 1 + e)$ and $r_2 \equiv p M /( 1 - e)$ for some $e$ and $p$, Vi\`{e}ta's formulas imply that the third solution must be $r_3 = 2 p M /( p-4)$. Vi\`{e}ta's formulas also imply that
\begin{subequations}
	\label{ELpe}
\begin{align}
	\left( E/\calM \right)^2 &= \frac{ (p-2)^2 -4e^2 }{ p(p-3-e^2) } , 
	\\
	\left( L/\calM \right)^2 &= \frac{ p^2 M^2 }{ p-3-e^2 },
\end{align}
\end{subequations}
which relates the geodesic parametrizations $(E/m,L/m)$ and $(e,p)$.

The subtlety here is that although $p$ and $e$ are motivated by the \textit{hope} that the radii of pericenter and apocenter are given by $r_- = r_1$ and $r_+ = r_2$, this is not guaranteed without additional restrictions: There is nothing in the discussion thus far which precludes $r_-$ or $r_+$ from being equal to $r_3$ instead of $r_1$ or $r_2$. We now characterize those portions of the parameter space where $r_\pm$ \textit{are} related to $e$ and $p$ in the expected way. 

Noting that the apocenter cannot lie below the pericenter, and that neither can lie below the event horizon, enforcing the expected identification between $r_1$, $r_2$, and $r_\pm$ implies that $r_2 \geq r_1 > 2M$. This can occur only when the eccentricity lies in the interval $e \in [0,1)$. Next, the radius of a geodesic must evolve freely between its pericenter and its apocenter, which occurs when $1 > (E/M)^2 > \Phi_\mathrm{eff}(r,m)$ for all $r \in (r_1,r_2)$. This and the eccentricity constraint together imply that the semi-latus rectum must satisfy $p \geq 6+2e$. These constraints are summarized by \eqref{separatrix}. They imply that $r_- = r_1$, $r_+ = r_2$, and $r_3 \leq r_- \leq r_+$.

In the space of geodesics parameterized by $e$ and $p$, the line $p = 6+2e$ is known as the separatrix. On it, $r_1 = r_3$ and $(E/M)^2$ is equal to the local maximum of $\Phi_\mathrm{eff}(\cdot,m)$, at least when $e>0$. Eccentric geodesics on the separatrix are homoclinic; they asymptotically approach an unstable circular orbit in both the future and the past. In fact, the homoclinic orbits have the same specific energies and specific angular momenta as the exactly-circular geodesics which they approach. However, while a homoclinic geodesic can be described by an eccentricity and a semi-latus rectum, an unstable circular geodesic cannot. This is because, for unstable circular orbits below $6M$, $r_+$ is equal to  $r_3$ instead of $r_2$.

\section{Quadrupole moment of a variable-length rod}
\label{App:quad}

Most of our discussion has characterized the relevant aspects of an object's internal structure by its quadrupole moment. However, no attempt has been made to connect those moments to particular internal models. This appendix provides some intuition by computing the quadrupole moment for two masses connected by a variable-length strut. For simplicity, we restrict to flat spacetime, in which case \eqref{Jint} can be used to compute $\tilde{J}^{abcd}$ in terms of a stress-energy tensor $T^{ab}$. We also assume that in the Minkowski coordinates $(t,x,y,z)$, the hypersurfaces $\Sigma_s$ are the hyperplanes with constant $t$.

Our first task is to find a stress-energy tensor which physically describes two masses connected by a variable-length strut. The main criteria are that $T^{ab}$ must (i) be conserved, (ii) have spatially-compact support, and (iii) have positive mass. As noted on page 89 of \cite{Wald}, conserved stress-energy tensors in flat spacetime can be written as
\begin{equation}
	T^{ab} = \nabla_c \nabla_d U^{acbd} ,
	\label{TU}
\end{equation}
where $U^{abcd}$ is constrained only to satisfy $U^{abcd} = U^{[ab]cd} = U^{ab[cd]} = U^{cdab}$. A slender body oriented along the $x$ axis might therefore be described by a $U^{abcd}$ whose only nonzero components follow from
\begin{equation}
	U^{txtx} (t,x,y,z) = m  \ell(t) U(x/\ell(t) ) \delta(y) \delta(z).
\end{equation}
Here, $m$ is a (constant) mass, $\ell >0$ a freely-specifiable length, and $U$ a dimensionless profile function. Substituting this into \eqref{TU} shows that the resulting stress-energy tensor has compact support when $U''(\bar{x})$ and $\bar{x} U'(\bar{x}) - U(\bar{x})$ vanish at large distances, say when $|\bar{x}| = | x/\ell | > 1$. In order to satisfy these conditions and to ensure that the mass is indeed given by $m$, it suffices to assume that
\begin{equation}
	U(\bar{x}) = \tfrac{1}{2} |\bar{x}|  
	\label{uBC}
\end{equation}
for all $|\bar{x}| > 1$. Next, the spatial components of the momentum vanish can be made to vanish, meaning that $p^a = m \partial_t$, when
\begin{equation}
	\int_{-\infty}^\infty \bar{x} U''(\bar{x}) d\bar{x} = 0.
	\label{uZeroP}
\end{equation}
This also ensures that $S^{ab} = 0$, so the center of mass condition \eqref{SSC} is satisfied.

In order to compute the corresponding quadrupole moment, it follows from \eqref{Jint} that its only nontrivial components are determined by
\begin{equation}
		\tilde{J}^{txtx} = \tfrac{3}{4} m \ell^2 \int_{-\infty}^{\infty} \bar{x}^2 U''(\bar{x}) d\bar{x} . 
\end{equation}
The trace-free quadrupole moment $J^{abcd} = (\tilde{J}^{abcd})_{\mathrm{TF}}$ therefore has the momentum component $\Pi^{ab} = 0$ and the mass component
\begin{equation}
	Q^{ab} = m \ell^2   ( \partial_x \otimes \partial_x )_\mathrm{TF} \int_{-\infty}^{\infty} \bar{x}^2 U''(\bar{x}) d\bar{x}.
	\label{Qint}
\end{equation}
Regardless of the time dependence of $\ell$, this is identical to the \textit{Newtonian} mass quadrupole which would result from a rod with mass density $(m/\ell) U''(x/\ell) \delta(y) \delta(z)$. 

As a more specific model, consider two identical point masses connected by a variable-length massless strut. Such a system can be described by
\begin{equation}
	U(\bar{x}) = \tfrac{1}{2} \left[ (\bar{x} + 1 ) \Theta(\bar{x} + 1 ) + (\bar{x} - 1 ) \Theta(\bar{x} - 1 ) - \bar{x} \right] ,
\end{equation}
where $\Theta$ denotes the Heaviside step function. The integral in \eqref{Qint} is then equal to unity. Moreover, the nonzero components of the stress-energy tensor are
\begin{align}
    T^{tt} = \tfrac{1}{2} (m/\ell)  \delta(y) \delta(z) [ \delta (\bar{x}-1) + \delta(\bar{x}+1)] ,
    \\
    T^{tx} = \tfrac{1}{2} (m/\ell) \delta(y) \delta(z)  [ \delta (\bar{x}-1) - \delta(\bar{x} + 1)] \partial_t  \ell,
    \\
    T^{xx} = \tfrac{1}{2} m \delta(y) \delta(z)  [\Theta (\bar{x} + 1) - \Theta (\bar{x}-1)] \partial_t^2 \ell 
    \nonumber
    \\
    ~ + T^{tt} (\partial_t \ell)^2.
\end{align}
Although there are first and second derivatives of $\ell$ in $T^{ab}$, no such derivatives appear in $J^{abcd}$. It may also be noted that if $d^2\ell/dt^2 \neq 0$, the interior of the strut carries a stress but has no energy or momentum density (as seen by an observer with 4-velocity $\partial_t$). The strut therefore violates the dominant energy condition\footnote{Energy conditions can be defined in an integrated sense for  distributional stress-energy tensors. For example, the dominant energy condition would correspond to demanding that  $\int T_{ab} \tau_1^a \tau_2^b dV \geq 0$ for all test vector fields $\tau_1^a$ and $\tau_2^a$ which are smooth, have compact support, and are future-directed timelike.}, as expected from its description as ``massless.'' Lastly, since $\ell$ is an arbitrary function of time, it is possible for the endmasses here to have spacelike trajectories. This occurs when $|\partial_t \ell| > 1$, in which case the dominant energy condition is violated not only in the interior of the rod, but also at its endpoints.

Using different choices for $U$, it is possible to construct models which do not violate energy conditions (at least when $\ell$ does not vary too rapidly). However, these models would no longer be interpreted as describing massless struts. Whether massless struts are retained or not, one may nevertheless  generalize the calculation in order to describe linkages which involve any number of masses and struts. Although such calculations might technically be performed assuming that the spacetime is flat, the resulting quadrupole moments would remain good approximations also in more general spacetimes, at least for bodies which are small compared to the local curvature scales.

\clearpage

\bibliography{refsLens}

\end{document}